\documentclass[12pt,letterpaper]{article}
\pdfoutput=1

\usepackage{graphicx,array}
\usepackage[dvipsnames]{xcolor}
\definecolor{darkblue}{rgb}{0,0.1,0.5}
\definecolor{darkgreen}{rgb}{0,0.5,0.2}
\definecolor{darkred}{RGB}{153,26,0}
\usepackage{ulem}
\definecolor{seablue}{rgb}{0,0.2,0.6}
\usepackage{latexsym}
\usepackage{amssymb, amsmath}
\usepackage{bm}        % for math
\usepackage[numbers,sort&compress]{natbib}
\usepackage{bm,relsize}
\usepackage{slashed}

\definecolor{viola}{RGB}{134,41,198}
\usepackage{mathrsfs}
\usepackage{hyperref} %Automatically links \label and \ref commands; Always load last
\hypersetup{
    colorlinks=true,       % false: boxed links; true: colored links
    linkcolor=darkblue,          % color of internal links
    citecolor=BrickRed,        % color of links to bibliography
    filecolor=magenta,      % color of file links
     urlcolor=MidnightBlue           % color of external links
}
\usepackage[all]{hypcap} %Link navagates to top of figure instead of caption (below fig)
\usepackage{subcaption}

\usepackage[font=small,labelfont=bf,labelsep=period]{caption}

\newcommand{\be}{\begin{equation}}
\newcommand{\ee}{\end{equation}}

\newcommand{\KeV}{\mathrm{KeV}}

 \date{\today}

\newcommand{\tq}{\mathtt{q}}

\newcommand{\Mpc}{\mathrm{Mpc}}
\newcommand{\cmqg}{\mathrm{cm}^2/\mathrm{g}}

\newcommand{\Trel}{T_{0,\mathrm{rel}}}

\begin{document}

%%%%%%%%%%%%%%%%%%%%%%%%%%%%%%%%%%%%%%%%%%%%%%%%%%%%%%%%%%%%%%%%%%%%%%%%%%
\begin{flushright}

\end{flushright}
\vspace{.6cm}
\begin{center}
{\huge \bf 
Dark matter self-interactions\\ \vspace{2mm}in the matter power spectrum
}\\
\bigskip\vspace{1cm}
{
\large Raghuveer Garani, Michele Redi, Andrea Tesi
}
\\[7mm]
 {\it \small
INFN Sezione di Firenze, Via G. Sansone 1, I-50019 Sesto Fiorentino, Italy\\
Department of Physics and Astronomy, University of Florence, Italy\\
 }

\end{center}
%%%%%%%%%%%%%%%%%%%%%%%%%%%%%%%%%%%%%%%%%%%%%%%%%%%%%%%%%%%%%%%%%%%%%%%%%%

\bigskip \bigskip \bigskip \bigskip

%%%%%%%%%%%%%%%%%%%%%%%%%%%%%%%%%%%%%%%%%%%%%%%%%%%%%%%%%%%%%%%%%%%%%%%%%%
\centerline{\bf Abstract} 
\begin{quote}
We study the imprints of secluded dark sectors with a mass gap and self-interactions on the matter power spectrum. When Dark Matter (DM) is sufficiently light, in the ballpark of a few KeV, and self-interacting we find qualitative difference with respect to $\Lambda$CDM and also to free streaming DM. In order to emphasize the role of interactions for the evolution of the primordial perturbations we discuss various regimes: ranging from the ideal case of a tightly coupled perfect fluid to the free case of Warm Dark Matter, including the realistic case of small but non-vanishing self-interactions. We compute the matter power spectrum in all these regimes with the aid of  Boltzmann solvers. Light dark sectors with self-interactions are efficiently constrained by Lyman-$\alpha$ data and we find that the presence of self-interactions relaxes the bound on the DM mass. As a concrete realization we study models with dark QCD-like sectors, where DM is made of light dark-pions.
\end{quote}
%%%%%%%%%%%%%%%%%%%%%%%%%%%%%%%%%%%%%%%%%%%%%%%%%%%%%%%%%%%%%%%%%%%%%%%%%%

%%%%%%%%%%%%%%%%%%%%%%%%%%%%%%%%%%%%%%%%%%%%%%%%%%%%%%%%%%%%%%%%%%%%%%%%%%
\vfill
\noindent\line(1,0){188}
{\scriptsize{ \\ E-mail:\texttt{  \href{garani@fi.infn.it}{garani@fi.infn.it}, \href{mailto:michele.redi@fi.infn.it}{michele.redi@fi.infn.it}, \href{andrea.tesi@fi.infn.it}{andrea.tesi@fi.infn.it}}}}
\newpage

\tableofcontents

\setcounter{footnote}{0}

%=========================================================================

\section{Introduction}

If Dark Matter (DM) belongs to a secluded dark sector one of the few hopes to study its properties relies on gravitational effects at cosmological scales. 
Secluded dark sectors nevertheless can be efficiently constrained, roughly speaking, if they deviate substantially from the predictions of $\Lambda$CDM through CMB or from observations of large scale structures through Lyman-$\alpha$ data~\cite{Viel:2013fqw,Tulin:2017ara,Palanque-Delabrouille:2019iyz,ParticleDataGroup:2020ssz,Garny:2018byk}, spanning a range of scales going from  giga to hundreds of parsecs. 

In this work we study self-interacting dark sectors with a mass gap, where DM is the lightest state with a mass in the KeV range.
 We assume that DM has elastic self-interactions, i.e. interactions that do not change the particle number, that keep the system in kinetic equilibrium while non-relativistic, effectively producing non-vanishing sound speed for the DM ``fluid''. Self-interactions and the lightness of DM  modify the evolution of DM perturbations. Here we quantitatively study the impact of DM self-interactions on the matter power spectrum, which provides the most stringent constraints.

The above scenario is realized  automatically in models motivated by accidental stability of DM. 
With minor exceptions if DM is accidentally stable the dark sector must contain new gauge interactions~\cite{Antipin:2015xia}. 
If the dark sector has no Standard Model (SM) charges it is also accidentally secluded and has non-trivial dynamics such as confinement and phase transitions.  Importantly for our purpose gauge interactions also equip DM with an unavoidable amount of self-interactions. One concrete realization of the above picture is provided by the dark-QCD sector studied in~\cite{Garani:2021zrr},
where DM is made of the lightest dark pions.

\begin{figure}[t]
\centering
\includegraphics[width=0.85\linewidth]{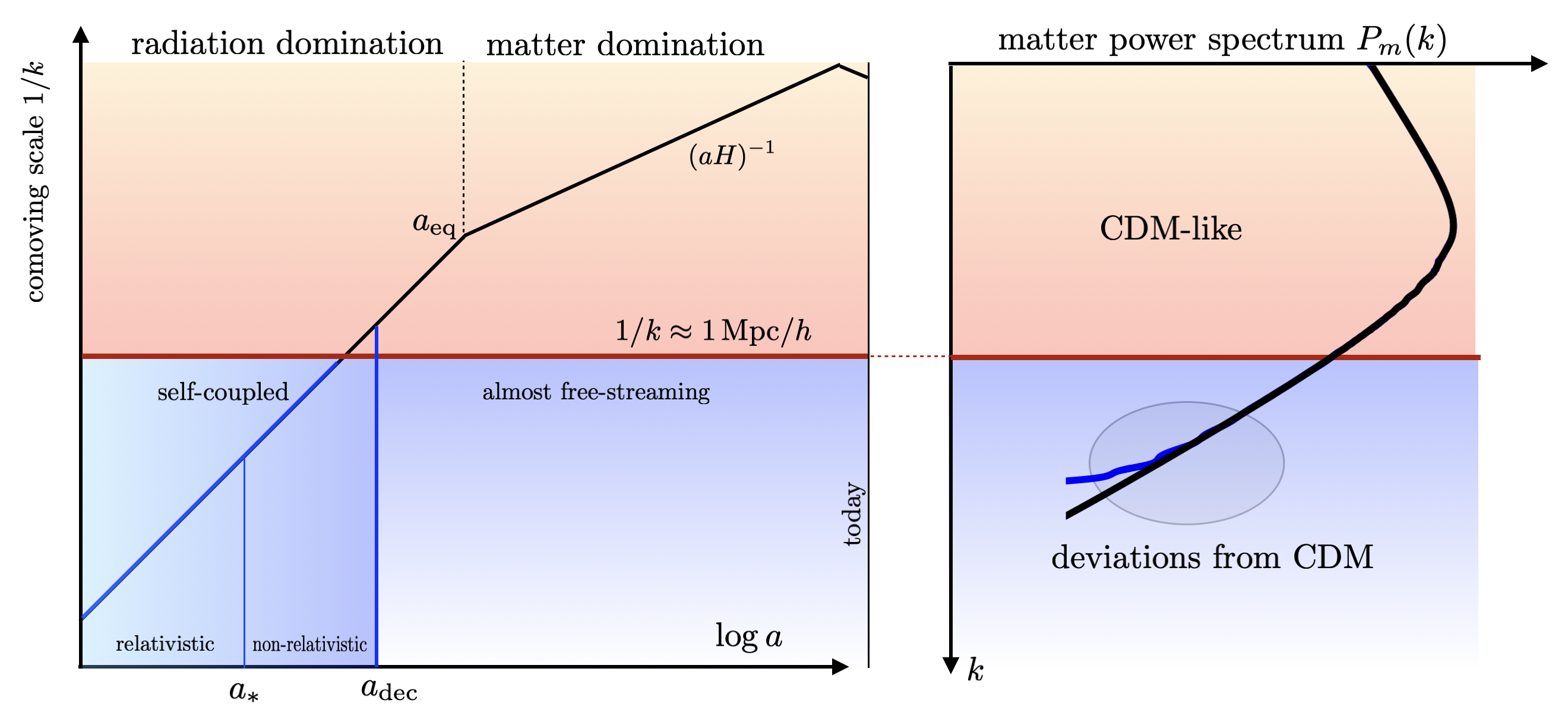}~
\caption{\label{fig:sketch}\it DM is produced at some early time during radiation domination and depending on self-interactions in the dark sector remains self-coupled in the
non-relativistic regime. For DM in KeV range this produces deviations from CDM in matter power spectrum for wave-numbers $k\sim h/{\rm Mpc}$ that are constrained
by  Lyman-$\alpha$ forest measurements.}
\end{figure}

Irrespective of the fundamental realization, given the DM mass $M$ and the elastic cross-section $\sigma_{\rm el}$,  physics can be summarized as in figure \ref{fig:sketch}. DM is produced with a given phase space distribution. If self-interactions are present the distribution quickly approaches an equilibrium distribution and DM particles remain self-coupled while becoming non-relativistic. This only requires very small cross-sections, orders of magnitude smaller than the bullet cluster constraint $\sigma_{\rm el}/M \lesssim 1\, \mathrm{cm}^2/\mathrm{g}$~\cite{Spergel:1999mh,Tulin:2013teo}. Depending on the size of the cross-section elastic interactions decouple deep in the non-relativistic regime and DM starts behaving as free-streaming warm DM. In light of currently available data sets, qualitative deviations from the behavior observed in the $\Lambda$CDM model are only possible at scales smaller than those probed by the CMB, corresponding to wave-numbers $k\gtrsim 1\, h/\Mpc$. On the contrary, at smaller scales deviations are possible, although one has still to confront with data, most notably from Lyman-$\alpha$ data, which give valuable information on the matter power spectrum at larger $k$ than the CMB~\cite{Viel:2013fqw}, see~\cite{Irsic:2017ixq} for an updated discussion. Also at galactic scales, DM self-interactions could be potentially constrained \cite{Salucci:2018hqu}. DM candidates that emerge from this picture are light and self-interacting, we dub this scenario Self-interacting Warm Dark Matter (SIWDM). 

We quantitatively study the scenario by computing the matter power spectrum with the  Boltzmann solver \texttt{CLASS}~\cite{CLASSII,CLASSIV}. 
This requires a modification of the publicly available code to include self-interactions, allowing DM to interpolate between perfect fluid and free streaming  behavior. Related work can be found in~\cite{Egana-Ugrinovic:2021gnu,Yunis:2021fgz,Erickcek:2021fsu}. While our analysis differs in several aspects our findings are consistent with the results presented in~\cite{Egana-Ugrinovic:2021gnu}.

Our results extend studies on Warm Dark Matter (WDM) widely studied in the literature~\cite{Hannestad:2000gt,Huo:2019bjf}. Non-interacting WDM can be produced from a thermal bath~\cite{Heeck:2017xbu,Bae:2017dpt,Boulebnane:2017fxw,Huo:2019bjf,Dvorkin:2020xga,DEramo:2020gpr} or via out-of-equilibrium decays and non-thermal mechanisms~\cite{Baldes:2020nuv,Decant:2021mhj,Baumholzer:2021heu,Dienes:2021cxp} with a given phase space distribution. This can affect structure formation if DM particles are sufficiently light and fast. We find that, adding DM self-interactions, even tiny, affects qualitatively the physics of  DM perturbations, and hence the matter power spectrum, reducing for example the constraint on the DM mass up to 20\%. 
We emphasize that DM in our context does not interact with any other (relativistic) component of the dark sector. For this reason, the scenario under consideration differs from models where DM is, for example, pressure-less and it interacts with a component of dark radiation possibly leading to dark-acoustic-oscillations (see~\cite{Buen-Abad:2015ova,Lesgourgues:2015wza,Buen-Abad:2017gxg,ETHOS,Huo:2017vef,Das:2018ons,Escudero:2018thh,Archidiacono:2019wdp}).

The paper is organized as follows. In section~\ref{sec:equations} we derive the equations that govern DM perturbations that we have implemented in \texttt{CLASS}, focusing on the new terms due to self-interactions. In section~\ref{sec:SIWDM} we determine the background  phase space distribution of DM imposing number and entropy conservation in the dark sector. In section~\ref{sec:power} we present numerical results for Weyl fermion: the matter power spectrum and the bound on DM mass obtained from the Lyman-$\alpha$ observations. In Section~\ref{sec:darkQCD} we apply our results to dark pion DM scenario introduced in~\cite{Garani:2021zrr}: a secluded QCD-like dark sector, where DM is in the form of dark-pions. We conclude in section~\ref{sec:conclusion}, while the appendix~\ref{app:CLASS} is dedicated to a discussion of our modifications of the code \texttt{CLASS}.

%%%%%%%%%%%%%%%%%%%%%%%%%%
%%%%%%%%%%%%%%%%%%%%%%%%%%

\section{Evolution of perturbations}\label{sec:equations}

In this section we derive the relevant perturbation equations for secluded DM that is produced at a given redshift, with a given phase space distribution (PSD). The distribution evolves such that the only interactions still active are of the elastic type (see figure~\ref{fig:sketch} for a cartoon). We numerically compute, with the aid of the code \texttt{CLASS}~\cite{CLASSII,CLASSIV}, the  matter power spectrum $P_m(k)$, defined as 
\be\label{eq:power-spectrum-definition}
\langle \delta(\vec{k},t_0)\delta(\vec{k}', t_0) \rangle = (2\pi)^3 \delta^{(3)}(\vec{k}+\vec{k}')P_m(k)\,,
\ee
 where $\delta(\vec{k},t_0)$ is the spatial Fourier transform of the density contrast $\delta\equiv \delta\rho/\rho$ of DM today, and $P_m(k)$ is measured in $[\Mpc/h]^3$. Below we highlight the approximations we have implemented in order to compute the latter.

In the absence of any other approximations schemes (such as the tight-coupling limit for the baryon-photon fluid in the visible sector), the density contrast must be computed from the phase-space distribution of DM, $f(\vec x, \vec p, t)$. The full distribution $f(\vec x, \vec p, t)$ is obtained by solving the associated Boltzmann equations that we now derive below. Here and in the following we adopt the convention
and notation of Ref.~\cite{Ma:1995ey} as implemented in \texttt{CLASS} \cite{CLASSII,CLASSIV}. We first briefly review the derivation of the Boltzmann equations
 following~\cite{Ma:1995ey} before we describe our approximations in the subsequent sections. 
 The DM distribution is defined from the homogeneous background distribution as
\be
f(\vec{x},\vec{q},\eta)=f_0(q,\eta)\left[1+\Psi(\vec{x},\vec{q},\eta)\right]\,,%\quad \longleftarrow \quad \frac{D f}{d\eta}=\left(\frac{\partial f}{\partial \eta}\right)_{\rm coll}\,,
\ee
where we have introduced the conformal time $dt=a d\eta$ and defined $\vec q \equiv \vec p /a$. Here and in the following we refer to $f_0$ as the solution of the homogeneous background Boltzmann equation, i.e. the phase space distribution of DM in absence of perturbations.

By following the standard procedure outlined in~\cite{Ma:1995ey}, the equation for the spatial Fourier transform (in $\vec k$-space) of the perturbation at linear order (in Newtonian gauge) reads
\be\label{eq:psi-general}
\dot\Psi + i \frac{q}{\epsilon} (\vec k \cdot \vec n) \Psi + \frac{d\log f_0}{d \log q} \left[ \dot\phi - i \frac{\epsilon}{q} (\vec k \cdot \vec n) \psi\right] =\frac{1}{f_0}\left(\frac{\partial f}{\partial \eta}\right)_{\rm coll}-\frac{d\log f_0}{d\eta}\Psi\,.
\ee 
where $\epsilon\equiv \sqrt{a^2M^2+q^2}$ is the co-moving energy of the DM particle, and $(\phi,\psi)$ are the two scalar perturbations of the FRW metric in Newtonian gauge. Where $\vec n=\vec q/q$ is the unit direction of the momentum of DM. The perturbations are a function of $(\vec k, q, \vec n, \eta)$, however we make the assumption that the only dependence on $\vec n$ is through the angle it forms with $\vec k$, i.e. we have a function $\Psi(k,q,\mu,\eta)$ where $\mu\equiv \hat k\cdot \vec n$. 

The above perturbation Eq.~\eqref{eq:psi-general} is in general difficult to solve, since it requires the knowledge of the collision term that would involve an integral over the unknown perturbations, and also due to the fact that $\Psi$ is function of several variables. The standard technique to make Eq.~\eqref{eq:psi-general} tractable is to  expand the perturbations in spherical harmonics through the Legendre polynomials $P_\ell(\mu)$, and by projecting out  the $\ell$-th moment, $\Psi\equiv \sum_{\ell} (-i)^\ell (2\ell+1) \Psi_\ell (k,q,\eta) P_\ell(\mu)\,$. With this the $\mu$-dependence is traded for an infinite tower of first order coupled differential equations\footnote{Notice that in Eq.~\eqref{eq:psi-general} there are only terms linear in $\mu=P_1(\mu)$ which are responsible for coupling different multipoles, since $\mu P_\ell(\mu)=(\ell+1)/(2\ell+1)P_{\ell+1}(\mu)+\ell/(2\ell+1)P_{\ell-1}(\mu)$.}
for the $\Psi_{\ell}$ usually referred to as the Boltzmann hierarchy
\begin{eqnarray}
\label{eq0}
\dot\Psi_0 +\frac{q}{\epsilon}k \Psi_1 + \dot\phi  \frac{d\log f_0}{d \log q} &=& -\frac{d\log f_0}{d\eta}\Psi_0 \,,\\
\label{eq1}
\dot\Psi_1 -\frac{q}{3\epsilon}k \left(\Psi_0 -2\Psi_2\right) + \frac{\epsilon}{3q} k \psi  \frac{d\log f_0}{d \log q} &=& -\frac{d\log f_0}{d\eta}\Psi_1 \,,\\
\label{eq2}
\dot\Psi_\ell -\frac{q}{(2\ell+1)\epsilon}k \left[ \ell \Psi_{\ell-1}- (\ell+1) \Psi_{\ell +1}\right] &=& \frac{1}{f_0} \left(\frac{\partial f}{\partial \eta}\right)_{\rm coll}^{(\ell)} -\frac{d\log f_0}{d\eta}\Psi_\ell\,,\quad  \ell\geq 2\,.
\end{eqnarray}
The initial conditions for this set of equations are assumed to be adiabatic.\footnote{They are related to integrated quantities such as the density contrast $\delta$ and velocity divergence $\theta$
\be
\Psi_0|_{(i)}=-\frac{1}{4}\frac{d\log f_0}{d \log q} \delta|_{(i)}\,,\quad \Psi_1|_{(i)}=-\frac{\epsilon}{3q k}\frac{d\log f_0}{d \log q} \theta|_{(i)}\,,\quad \Psi_2|_{(i)}=-\frac{1}{4}\frac{d\log f_0}{d \log q} \sigma|_{(i)}\,,\quad \Psi_{\ell\geq 3}|_{(i)}=0\,,
\ee
where $\sigma$ is the anisotropic stress perturbation. Notice that they only depend on $q$ through $f_0$.
}

The effect of DM self-interactions enter the evolution equations through the moments $\ell\geq 2$. While the effect of the eventual time dependence of the background solution $f_0$ is captured by a universal coefficient given by $d\log f_0/d\log \eta$ (for a similar structure of the equations see also~\cite{Chacko:2019nej}) for all the moments.\footnote{The collisional term vanishes for $\ell=0,1$ due to co-moving number density and energy conservation, respectively.}
Note that the gauge dependencies appear only in Eqs.~\eqref{eq0} and~\eqref{eq1}. In order to numerically solve the hierarchy equations, we have to truncate the  above to some large moment $\ell_{\rm max}$. To this end we follow the truncation method introduced in~\cite{Ma:1995ey} as implemented in \texttt{CLASS} (see appendix~\ref{app:CLASS} for more details). The above set of equations governs the evolution of the perturbation of DM in full generality. To make further progress some approximations are needed.

\subsection{The role of interactions}\label{sec:interactions}
As mentioned before, the collision terms related to self-interactions only contribute to multipoles larger than two ($\ell >2$). Physically, the role of the collision terms is to suppress perturbations on higher multipoles. Using the so-called relaxation time approximation the collisional terms simplify to
\be
\frac{1}{f_0} \left(\frac{\partial f}{\partial \eta}\right)_{\rm coll}^{(\ell)} \approx - \frac{1}{\tau_{\rm coll}} \Psi_\ell\,,
\ee
where the (co-moving) relaxation time is defined by $\tau_{\rm coll}$. Notice that this structure of the interaction terms is analogous to the visible sector, i.e. when considering higher multipoles for perturbations of the photon, where collision time is set by the Thomson scattering.
Given an elastic cross-section $\sigma_{\rm el}$, the co-moving relaxation time can be computed as
\be\label{eq:taucoll}
\tau_{\rm coll}\equiv\frac{a^{-1}}{n(a)  \sigma_{\rm el} v} \approx 1.6\times 10^{5} \Mpc\, \left(\frac{0.12}{\Omega h^2}\right)\left(\frac{1\, \cmqg}{\sigma_{\rm el}/M}\right)\times \frac{a^2}{v(a)}\,.
\ee
Here $v(a) = 1/\sqrt{3}\,a_\star/a $ in the non-relativistic limit, and  $a_\star$ denotes the scale factor when DM effectively becomes non-relativistic (see section~\ref{sec:psd} for further details). This approximation is often used to simplify otherwise intractable collisional terms (see for example~\cite{Egana-Ugrinovic:2021gnu,Yunis:2020woq}).
It is instructive to analyze first two limiting cases: the limit where the collisions are irrelevant ($\tau_{\rm coll}\to \infty$) and when they are the fastest scale in the problem ($\tau_{\rm coll}\to 0$). The first limit corresponds to the limit where DM is truly collisionless, while the latter corresponds to the case where DM behaves essentially as a perfect fluid. Finally, note that the form of the relaxation time slightly differes from that of ref.~\cite{Hannestad:2000gt} due to contact interactions we consider. 

In this work we only consider the case where $\sigma_{\rm el}$ is constant in time, i.e. without any appreciable energy and velocity dependence. Approximately, DM particles will be in kinetic equilibrium  if $ a H \tau_{\rm coll}\ll 1$. From this condition we estimate the scale factor at decoupling to be
\be
\label{eq:adec}
a_{\rm dec}\approx 5 \times 10^{-4}
%\left(\frac{0.67}{h}\right)^{1/2}
%\left(\frac{0.126}{\alpha}\right)^{1/2}
\left(\frac{\Omega h^2}{0.12}\right)^{2/3}
\left(\frac{\sigma_{\rm el}/M}{1\, \mathrm{cm}^2/\mathrm{g}}\right)^{1/2}
\left(\frac{\rm KeV}{M}\right)^{2/3}\,.
\ee
Note that the bullet cluster constraint on self-interaction requires $\sigma_{\rm el}/M \lesssim \rm{cm^2/g}$ corresponding $a_{\rm dec}\sim a_{\rm eq}$ for masses in the KeV range. Nevertheless, note that even for much smaller cross-sections kinetic decoupling takes place in the non-relativistic regime.

\subsubsection{Strong interactions: perfect fluid}
When the self interacting cross section is large, $\tau_{\rm coll}$ is the shortest time scale in the problem and consequently all the higher multipoles relaxes quickly to zero, i.e. $\Psi_{\ell\geq2}= 0$. Thus the perturbation equations collapse to only two equations~\eqref{eq0} and~\eqref{eq1} which correspond to density and pressure perturbations, respectively. They can be integrated over the DM momentum $q$, with appropriate weights,\footnote{We recall that the perturbations are
$\delta=\int d^3q\, \epsilon\, f_0\,\Psi_0/(\rho a^4)$ and $\theta=k \int d^3q\, q\, f_0 \Psi_1/(\rho(1+w) a^4)$.}
 to recover the expressions for $\delta$ and for the velocity divergence $\theta$, which are the only two variables relevant for perfect fluids. The evolution equations read
\begin{eqnarray}\label{perfect1}
\dot \delta + (1+w)\theta+3H_\eta \delta \left(c_s^2-w\right ) &=&3(1+w)\dot\phi\,,\\
\label{perfect2}
\dot \theta + H_\eta (1-3w)\theta +\frac{\dot w}{1+w} \theta -\frac{c_s^2}{1+w} k^2\delta &=& k^2\psi\,,
\end{eqnarray}
where $H_\eta\equiv \dot a(\eta)/a(\eta)$ and $w=p/\rho$ is the equation of state of the fluid. Notice that the above equations hold independently of the explicit form of $f_0$. The equation of state $w=p/\rho$ and the sound speed $c_s^2$ are related by the differential equation $c_s^2 =\dot p/\dot\rho$. In the relativistic and non-relativistic regime one finds respectively,
\be\label{eq:perfect-matching}
\mathrm{rel:}\quad c_s^2=w=\frac{1}{3},\quad\quad \mathrm{non-rel:}\quad c_s^2=\frac{5}{3}w=\frac{1}{3}\frac{a_*^2}{a^2}\,,
\ee
where $a_*$ can be obtained, for example, by matching to the non-relativistic regime of $f_0$, see section~\ref{sec:SIWDM}. A smooth interpolation between these two regimes can be achieved by computing at each time $c_s^2=\dot p/\dot \rho$.

It is instructive to rewrite the above equations by making a few simplifications in order to highlight the behavior we expect in the matter density contrast for the realistic case of relatively large $k$ (for small wave-numbers, instead, DM must be well non-relativistic). For large $k$, where DM can be still semi-relativistic, the two equations can be combined and rewritten as a second order differential equation for the density contrast. In the relevant regime $a_* H(a_*)\gg k\gg a_{\rm eq}H(a_{\rm eq})$
\be\label{eq:perfect-approx}
\ddot\delta + H_\eta \dot\delta + c_s^2(a) k^2 \delta \approx -k^2\psi, \quad\quad \mathrm{for\, large}\ k\,,
\ee
where the dependence of $w,\,c_s^2$ on the scale factor is given by the right part of Eq.~\eqref{eq:perfect-matching}. The sub-horizon evolution of these modes do not display sizable `dark acoustic oscillations' since the the sound horizon is shrunk by $c_s^2\sim 1/a^2$.  Therefore, the relevant parameter for a perfect fluid is the Jeans' scale related to the sound horizon
\be
k_J(a) \equiv a H(a)/c_s(a)\,.
\ee
On the contrary, for modes with $k\gg a_*H(a_*)$ we should expect the imprint of truly `dark acoustic oscillations'  onto the power spectrum, since for these modes the sound speed is the one of a relativistic fluid until $a\sim a_*$. The oscillations will have a `period' of order $\Delta k \sim \mathrm{a\ few} \times  a_* H(a_*)/c_s$. As we will discuss later, these $k$-modes correspond to wave-numbers at the edge or beyond the Lyman-$\alpha$ range. We stress however that this situation is qualitatively different from the `dark acoustic oscillations' discussed in the context of DM interacting with dark radiation, for example in~\cite{ETHOS}, where the presence of a tight-coupling between DM and dark radiation can generate oscillations at smaller scales.
\subsubsection{Weak interactions: CDM and free-streaming}
Without interactions the collisional term on the right hand side of the Eqs.~\eqref{eq2} vanish (corresponding to $\tau_{\rm coll}\to \infty$). In this regime the system of Eqs.~\eqref{eq0}-\eqref{eq2} simplify, but it can be still difficult to solve. To examine this case further it is convenient to rewrite the hierarchy equations in a dimensionless form (for all $\ell$) as
\be
\frac{d\Psi_\ell}{d\log a} =\frac{1}{a H}\frac{q}{\epsilon}\, k \, M_{\ell,\ell'}\Psi_{\ell'}  -\frac{1}{a H}\frac{d\log f_0}{d\log q}\times\big[a H \frac{d\phi}{d\log a} \delta_{l,0} + \frac{\epsilon}{3q} k \psi \delta_{\ell,1}\big]\,,
\ee
where $M_{\ell,\ell'}\equiv \big[\ell\delta_{\ell-1,\ell'}-(\ell+1)\delta_{\ell+1,\ell'}\big]/(2\ell+1)$ is a dimensionless matrix.
In this dimensionless form it is manifest that the evolution of the moments $\Psi_\ell$ depends on the size of $k$ as compared to a parameter that appears in the above equation
\be\label{eq:kfs}
k_{\rm fs}\equiv a H \, \epsilon/q\equiv a H\, /u(a)\,,
\ee
with $u\equiv q/\epsilon$, a co-moving velocity. This quantity scales as $\sim 1/a$ when non-relativistic and thus decreases with the scale factor. Further note that $ k_{\rm fs}$ depends on time and on $q$. This parameter allows us to understand the behavior of the perturbations. Qualitatively we can have two different regimes depending on the relative size of $k$ and $k_{\rm fs}$.

\paragraph{CDM}~\\
For $k<k_{\rm fs}$, higher moments are suppressed by $(k/k_{\rm fs})^\ell$ and the system can be truncated consistently to the first two equations at $O(k/k_{\rm fs})$, leading to
\be
\dot\Psi_0 +\frac{q}{\epsilon}k \Psi_1 + \dot\phi  \frac{d\log f_0}{d \log q} = 0\,,\quad \dot\Psi_1+ \frac{\epsilon}{3q} k \psi  \frac{d\log f_0}{d \log q} =0\,.
\ee
By integrating the above equations over the DM momentum $q$ with appropriate weights we obtain the continuity and Euler equations of CDM (collisionless cold ``fluid'' with $c_s=w=0$), which is a species for which $k$ is always smaller than $k_{\rm fs}(a)$, for all the relevant modes. It never free-streams despite having no interactions.

\paragraph{Free-streaming DM}~\\
For $k>k_{\rm fs}$, all the higher moments are important. And for sub-horizon evolution ($k\eta\gg 1$), where scalar potentials are often negligible, one can show that the system of equations approximate to $\dot\Psi_\ell \approx M_{\ell,\ell'} \Psi_{\ell'}$. Since $M_{\ell,\ell'}$ is a tri-diagonal matrix with imaginary eigenvalues, the solution is a set of oscillators that suppresses the growth of the dark matter density contrast. The solutions oscillate as $\propto \exp(i k\int^\eta u(\eta) \eta )$.\footnote{Notice the absence the factor of $1/\sqrt{3}$ inside the argument of the sine: differently from the perfect fluid case these oscillations will have a slightly shorter period.} This is the usual effect of free-streaming ``warm'' dark matter, which can be semi-relativistic in the early stages of its evolution. It is common to parameterize this effect by computing a physical length today defined as \cite{Kolb:1990vq}
\be\label{eq:free-streaming}
\lambda_{\rm FS}\equiv\int^{\rm today} d\eta\, \langle v \rangle\,,\quad\quad \mathrm{free-streaming\ length}\,.
\ee

\paragraph{The effect of interactions on the higher multipoles}~\\
In order to emphasize another difference with respect to the case of perfect fluid, here we also expand the hierarchy of multipoles at first order in $1/\tau_{\rm coll.}$, retaining therefore the first correction from an ideal perfect fluid case. Notice that this expansion is only meaningful up to $\tau_{\rm coll.}\lesssim 1/H_\eta$. In doing so, we can close the system at $\ell=2$ and obtain from Eq.~\eqref{eq2},
\be
\Psi_2 = \frac{2}{5}\frac{q}{\epsilon}k \tau_{\rm coll} \Psi_1 +\cdots~.
\ee
This introduces a new scale in the equation analogous to the Silk-damping. We dub this scale as
\be\label{eq:damping}
k_{\rm coll}\equiv \sqrt{a H(a)/\tau_{\rm coll}}\,  \epsilon/q\,.
\ee
This quantity determines on what scales interactions give an exponential suppression to the density contrast (more properly to $\Psi_0$). This effect is ascribed to diffusion and it can be active on wave-numbers smaller than those corresponding to free-streaming Eq.~\eqref{eq:kfs}, therefore leading to a further suppression of power. In concrete realizations this new term does not significantly modify the predictions for the matter power spectrum. The corrections due to this term are $< \mathcal{O}(1)$. These comments directly apply to us since we will discuss scenarios where DM, in order to comply with observations, is not relativistic at matter-radiation equality. This is different from the case of photon perturbations for which $q=\epsilon$ and the effect is always important.

%%%%%%%%%%%%%%%%%%%%%%%%%%%%%%%
%%%%%%%%%%%%%%%%%%%%%%%%%%%%%%%

\section{Self-interacting warm DM}\label{sec:SIWDM}

Imprints of collision-less WDM on the matter power spectrum have been discussed at length in the literature as an example of free-streaming DM candidate that is produced semi-relativistically. After production, evolution of DM perturbations depends on the background distribution function $f_0(q)$ that is fixed by the production mechanism. Example of such non cold relics have been studied in the context of DM produced through decays and scatterings~\cite{Heeck:2017xbu,Bae:2017dpt,Boulebnane:2017fxw,Huo:2019bjf,Dvorkin:2020xga,DEramo:2020gpr}, or through evaporation of primordial blackholes~\cite{Decant:2021mhj}, and sterile neutrino DM~\cite{Viel:2006kd,Boyarsky:2008xj}.

 In this paper we discuss the phenomenology of WDM with the inclusion of self-interactions in the dark sector.
This scenario is motivated by secluded sectors obtained from gauge theories \cite{Hambye:2009fg,Hambye:2008bq,Garani:2021zrr} where self interactions are unavoidable (see section \ref{sec:darkQCD}). Moreover this scenario appears to be rather generic as it just requires very small self interactions. From Eq.~\eqref{eq:adec} we find that DM remains self-coupled in the non-relativistic regime as long as
\begin{equation}
\sigma_{\rm el} >\frac {10^{-20}}{\rm KeV^2} \left(\frac { M}{{\rm KeV}}\right)^{1/3}= 10^{-36}\,{\rm cm^2} \left(\frac { M}{{\rm KeV}}\right)^{1/3}\,.
\label{eq:sigmaelmin}
\end{equation}
This tiny cross-section is completely negligible for the dynamics of DM today but can have observable effects for structure formation.
Depending on the strength of the interactions the system interpolates between the asymptotic configurations discussed in the previous sections: 
when $\sigma_{\rm el}\sim {\rm cm^2/g}$ SIWDM will behave like a perfect fluid or as free-streaming DM for cross-sections smaller than 
(\ref{eq:sigmaelmin}).
 
When WDM has interactions the growth of perturbations is affected in two ways. First it modifies the background distribution function $f_0(q,\eta)$ 
that, contrary to free streaming DM, becomes a function of time and second it modifies the evolution of perturbations as discussed in the previous section.

\subsection{Background distribution function $f_0$}\label{sec:psd}

To study the evolution of cosmological perturbation the first step is to determine the background distribution $f_0(q,\eta)$ (in freely-available Boltzmann solvers such as \texttt{CLASS} \cite{CLASSII} 
it is indeed one of the inputs, see especially \cite{CLASSIV}).
We assume that number changing processes are not active so that the background distribution is in general determined by elastic scatterings through the (homogeneous) Boltzmann equation,
\begin{equation}
\begin{split}
\left(\frac{\partial f_0}{\partial \eta}\right)_{\rm coll}\equiv\frac{\partial f_0(q,\eta)}{\partial \eta} - H_\eta \frac{\partial f_0(q,\eta)}{\partial\log q}=\frac 1{2\epsilon(q,\eta)}\int \Pi_{i=1}^3 \frac {d^3 q_i}{(2\pi)^3 2E_i} (2\pi)^4\delta^{(4)}(Q)  |\mathcal{M}_{\rm elastic}|^2\\
\times \big[f_0(q_1)f_0(q_2)(1\pm f_0(q_3))(1\pm f_0(q)) -f_0(q_3)f_0(q)(1\pm f_0(q_1))(1\pm f_0(q_2))\big] \,.
\end{split}
\end{equation}

When the rate of elastic interactions is faster than Hubble, DM is in kinetic equilibrium, 
$f_0$ approaches the equilibrium distribution
\be\label{eq:f0}
f_0(q,\eta)=g_s \left[\exp{\left(\frac{E(q)-\mu}{T_D}\right)}\pm 1\right]^{-1}\,,
\ee
where $E(q)=\sqrt{q^2/a^2+M^2}$ and $T_D$ and $\mu$ are the DM temperature and chemical potential. If SIWDM remains kinetically coupled while becoming non-relativistic, $f_0$ will freeze out as a Maxwell-Boltzmann distribution, else as a Fermi-Dirac (Bose-Einstein) distribution. Note that the distribution is constant in the relativistic and non-relativistic regime regardless of interactions. The only time dependence arises in the transition between the two regimes  if this takes place in kinetic equilibrium. In light of Eq.~(\ref{eq:sigmaelmin}) this requires a very small cross-section.

When the system is in kinetic equilibrium number of particles and entropy per co-moving volume are conserved\footnote{Our formula become exact when the rate of elastic interactions is much larger than the expansion of the universe during the transition to non-relativistic regime, $H/\Gamma_{el}\to 0$. In light of eq. (\ref{eq:sigmaelmin}) this is an excellent approximation in a wide range of parameter space.}. These can be expressed in terms
of distribution functions as~\cite{bernstein_1988},
\begin{itemize}
\item \textbf{Number conservation}. 
\be\label{eq:number-conservation}
n a^3= \frac{1}{(2\pi)^3} \int  d^3q f_0(q,t)\,
\ee
\item \textbf{Entropy conservation}.
\begin{equation}\label{eq:entropy-conservation}
sa^3 \equiv \frac {(\rho+p -\mu n)}{T} a^3=-\frac{1}{(2\pi)^3} \int  d^3q   \bigg[ f_0 \log f_0 \pm (1\mp f_0) \log(1\mp f_0) \bigg ]
\end{equation}
\end{itemize}

Since the background distribution depends only on $T_D$ and $\mu$, the above two conditions are  sufficient to determine the temperature and the chemical potential as a function of time. 

\begin{figure}[t]
\centering
\includegraphics[width=0.5\linewidth, height=0.35\linewidth]{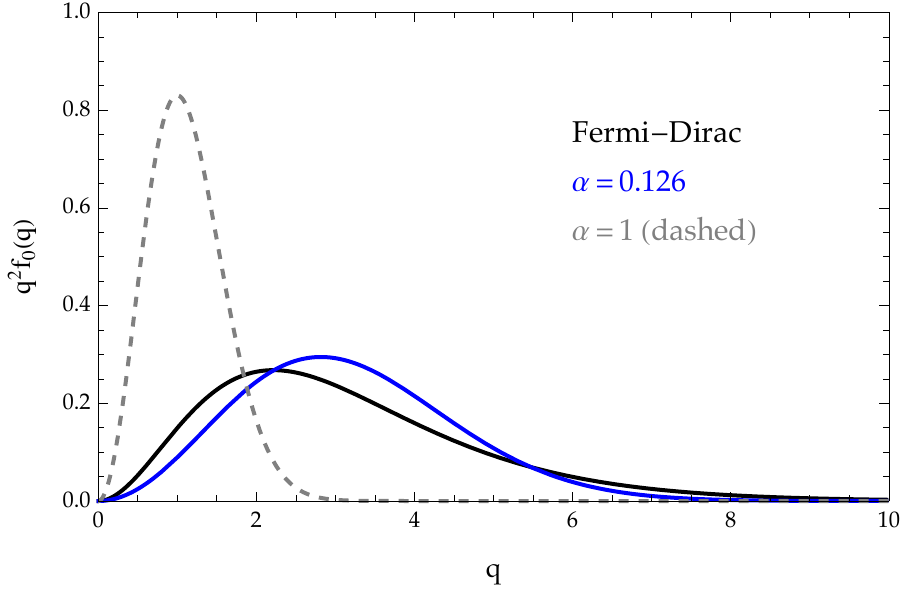}~
\includegraphics[width=0.5\linewidth,height=0.37\linewidth]{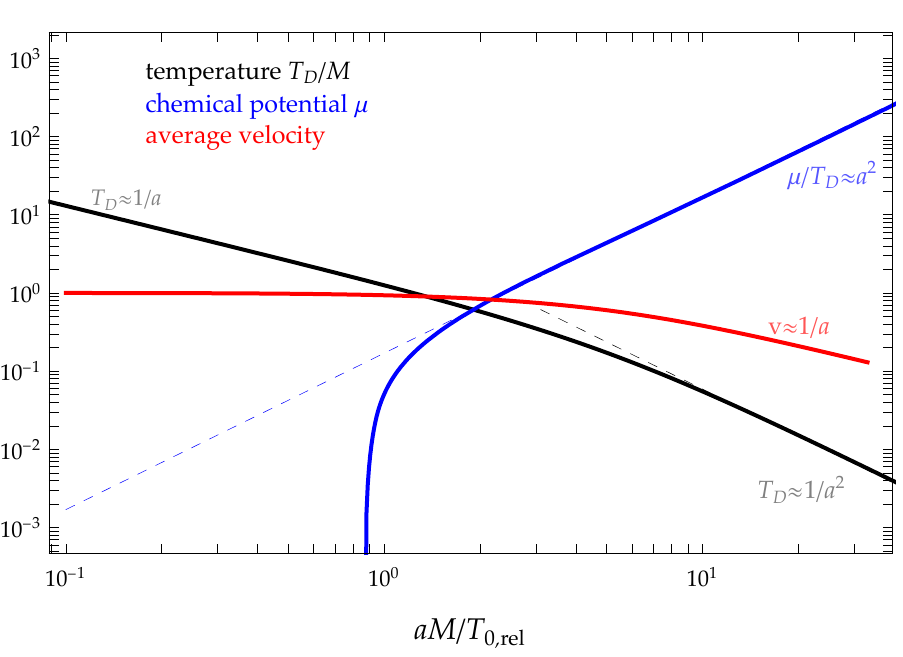}
\caption{\label{fig:comparef0}\it Left: Distribution functions $\tq^2 f_0(\tq)$ normalized to unity. Right: Temperature $T_D/M$ and chemical potential $\mu$ as a function of the scale factor when DM is self-coupled through elastic scatterings. The two functions allows a complete determination of the background distribution function $f_0(q,T_D,\mu)$ at all times while the system remains self-coupled crossing the relativistic/non-relativistic transition. }
\end{figure}

\subsection{Matching from relativistic to non-relativistic distributions}

The functions $T_D(a)$ and $\mu(a)$ can be in principle determined numerically inverting the above two Eqs.~\eqref{eq:number-conservation} and~\eqref{eq:entropy-conservation}.
As we will explain the evolution of perturbations relevant cosmologically is however insensitive to the details of the transition region.
Because of this, in practice, we will only need precisely the parameters of the non-relativistic distribution that can be used as an input to numerical simulations. Below we now show this can be obtained analytically by imposing number and entropy conservation away from the transition region. 

\paragraph{Relativistic regime}~\\
In this case of interest, during the initial stages of the evolution, DM is relativistic. The temperature appearing in $f_0$ is
\be
T_D=T_{0,\rm rel}/a\,.
\ee
Where $T_{0,\rm rel}$ is the temperature of DM today if it self-decoupled while relativistic. It is the temperature today if DM were still relativistic (i.e., it has the same meaning of the neutrino temperature in the SM). Therefore we have from Eq.~\eqref{eq:f0}
\be\label{f0rel}
f_0^{\mathrm{rel}}=\left[\exp{\left(\tq-\mu_0\right)}\pm 1\right]^{-1}\,.
\ee
Where we have defined  $\tq  \equiv q/T_{0,\rm rel}$, the dimensionless momentum variable used  in \texttt{CLASS}. 

Note we assume no initial chemical potential in the following.
From Eq.~\eqref{f0rel} we derive the DM abundance and yield to be 
\begin{equation}
\Omega h^2\approx 0.12 \frac{Y_{\rm DM} M_{\rm DM}}{0.44\,{\rm eV}} \,,~~~~~~~~~~~~ Y_{\rm DM}= g_D \frac{45 \zeta(3)}{2\pi^4 g_*^s(T_{0,\rm CMB})}  \left(\frac{T_{0,\rm rel}}{T_{0,\rm CMB}}\right)^3\,.
\label{eq:YDM}
\end{equation}
Also note that the above expression applies only for the case where WDM is produced in thermal equilibrium.
Through the abundance we determine the DM temperature as 
\be
\mathtt{T_{0,\rm rel}}=\frac{T_{0,\rm rel}}{T_{0,\rm CMB}}=0.16  \left(\frac{\mathrm{KeV}}{M}\right)^{1/3}  \left(\frac{\Omega h^2}{0.12}\right)^{1/3}\left(\frac{\mathrm{3/2}}{g_D}\right)^{1/3}\,, 
\label{eq:T0rel}
\ee
where $g_D=3/2\, (N)$ for a Weyl fermion (bosons) and $g_*^s(T_{0,\rm CMB})=3.94$ and $T_{0,\rm CMB}=2.348\times 10^{-4}\, \mathrm{eV}$ is the CMB photon temperature today, respectively. This input quantity sets the temperature of dark sector in \texttt{CLASS}, and is denoted by $\mathtt{T_{ncdm}}=\mathtt{T_{0,\rm rel}}$.

\paragraph{Non-relativistic regime}~\\
If DM  remains self-coupled when the temperature drops below its mass, $f_0$ evolves into a non-relativistic distribution. 
In this regime the dark sector temperature scales as $T_D\sim 1/a^2$. We define $T_{0,\mathrm{NR}}$ to be the temperature that DM has today if it self-decoupled in this regime. 
Expanding $E\approx M +q^2/(2M a^2)$ we have,
\be\label{f0NR}
f_0^{\mathrm{NR}}= e^{\mu_0^{\rm NR}}\exp{(-\alpha \tq^2)}\,,\quad\quad \alpha \equiv\frac{(T_{0,\rm rel})^2}{2M T_{0,\mathrm{NR}}}\,.
\ee

\paragraph{Matching}~\\
The parameter $\alpha$ and the chemical potential $\mu_0^{\rm NR}$ can be easily extracted by computing  number of particles and entropy
in the relativistic and non-relativistic regimes\footnote{Note that this is different from simply red-shifting Maxwell-Boltzmann distribution as done in~\cite{Trautner:2016ias}.}. One finds ($s= (\rho+p -\mu n)/T$),
\begin{eqnarray}
&&n_{\rm NR}= e^{\mu_0^{\rm NR}}\left(\frac {M T_D}{2\pi}\right)^{3/2}\,,~~~~~~\rho_{\rm NR}= M n_{\rm NR} +\frac 3 2 n_{\rm NR} T_D\,,~~~~~~~~p_{\rm NR}= n_{\rm NR} T_D \nonumber \\
&&n_{\rm REL}= g_D' \frac{\zeta(3)}{\pi^2} T_D^3\,,~~~~~~\rho_{\rm REL}=3p_{\rm REL}=g_D\frac{\pi^2}{30} T_D^4
\end{eqnarray}
where $g_D$ is the number of degrees of freedom for bosons and 7/8 for fermions  and $g_D'=1\,(3/4)$ per boson (fermions), respectively. Imposing conservation of number and entropy per co-moving volume, using Eqs.~\eqref{eq:number-conservation} and~\eqref{eq:entropy-conservation}, one 
can extract the all important parameter $\alpha$ of the non-relativistic  distribution in Eq.~\eqref{f0NR}. 

For fermions we find
\be
\label{eq:alphaF}
\alpha\big|_{\rm fermion}=\frac{\pi^{1/3} e^{\frac{5}{3}-\frac{14 \pi ^4}{405 \zeta (3)}}}{(6\zeta(3))^{2/3}} = 0.126164\,,
\ee
while for bosons,
\be
\label{eq:alphaB}
\alpha\big|_{\rm boson}=\frac{\pi^{1/3} e^{\frac{5}{3}-\frac{4 \pi ^4}{135 \zeta (3)}}}{4 \zeta(3)^{2/3}} = 0.155395\,.
\ee
Similarly the chemical potential can be derived.
The difference between bosons and fermions arises from a factor of $3/4$ and $7/8$ in the number and entropy densities respectively. 

We emphasize that these results are exact if the system evolves from relativistic to non-relativistic in kinetic equilibrium. As can be seen from the left panel in Fig.\,(\ref{fig:comparef0}) the distributions in the non-relativistic and the relativistic regimes have very similar shapes. This is intuitive as conservation of entropy demands that the distribution should changes as little as possible.

Furthermore, another important variable is the averaged squared velocity of DM. This can be computed by weighting the background distribution with $v^2=q^2/(a M)^2$. As we are not imposing energy conservation while transitioning from relativistic to non-relativistic regimes (differently from~\cite{Egana-Ugrinovic:2021gnu}), $\langle v^2\rangle$ varies. Depending on whether DM decouples in the relativistic or non-relativistic regime, the square velocity  for fermions is given by
\be\label{eq:velocity}
a^2 \langle v^2 \rangle|_{f_0^{\rm rel}}= \frac{15 \zeta(5)}{\zeta(3)}\frac{T_{0,\rm rel}^2}{M^2}\approx 12.9 \frac{T_{0,\rm rel}^2}{M^2} \,,
\quad a^2\langle v^2 \rangle|_{f_0^{\rm NR}}=\frac{3}{2\alpha}\frac{T_{0,\rm rel}^2}{M^2}\approx 11.9 \frac{T_{0,\rm rel}^2}{M^2}\,.
\ee
Similarly, for bosons one finds that the velocity is reduced by a factor $\mathcal{O}(10\%)$ for non-relativistic decoupling. 
The reduced velocity suggests that the constraints on interacting warm  DM will be weaker compared to free streaming.
This effect however account only for about 100 eV reduction of the mass. The main difference arises from the fact that 
DM behaves as a perfect fluid well into the NR regime (see section~\ref{sec:power}).

The coefficient $\alpha$ also determines the scale factor ($a_\star$) where DM becomes effectively non-relativistic (we have already introduced the parameter $a_*$ for the  perfect fluid sound speed in Eq.~\eqref{eq:perfect-matching}). This parameter can be used to obtain the averaged velocity squared. Which in the non-relativistic limit is given by $\langle v^2\rangle=(3/5) a_*^2/a^2$ (and for reference $c_s^2=(1/3) a_*^2/a^2$). Upon inspection of Eq.~\eqref{eq:velocity} we have
\be\label{eq:astar}
a_* = \sqrt{\frac{5}{2\alpha}} \frac{\Trel}{M}=\frac {5.95} {\sqrt{\alpha}} \times10^{-8} \left(\frac{\mathrm{KeV}}{M}\right)^{4/3}  \left(\frac{\Omega h^2}{0.12}\right)^{1/3} \left(\frac{\mathrm{3/2}}{g_D}\right)^{1/3}\,.
\ee
The above value of the scale factor is the input for our numerical simulations, it is used to discriminate with a step function between the relativistic and non-relativistic regimes. This formula also explicitly shows that if there are more than one degenerate species, $N>1$, the average velocity drops as $\langle v^2 \rangle \sim 1/N^{2/3}$, for fixed mass, making evident that at large $N$ smaller DM masses will be allowed.

While not crucial for our results an approximate solution for the transient can be found as follows.
Neglecting the quantum statistics, that is anyway irrelevant at low velocities we can parametrize the distribution function in kinetic equilibrium as,
\begin{equation}
f_0(q,\eta)\sim A  \exp{\left(- \frac{E(q)}{T_D}\right)}
\label{eq:f0approx}
\end{equation}
With this parametrization by imposing particle number conservation we can determine $A=2\pi^2 n a^3/\int dq q^2 e^{-E(q)/T}$. Imposing conservation of entropy we can then determine numerically $T(a)$.
The resulting functions (for a Weyl fermion) are shown in right panel of figure \ref{fig:comparef0}. From the numerical solution we see that the transient takes places when $a M/T_{0,\rm rel}\approx O(1)$, and we are already well into the non-relativistic phase if $a M/T_{0,\rm rel}\approx O(10)$. For practical purposes we approximate distribution function at all times as
$f_0(q, a)|_{\rm ansatz} = S(a M/\Trel) f_0^{\rm rel}(q)+ [1-S(a M/\Trel)] f_{0}^{\rm NR}(q).$
We have compared the results of the moments of the real distribution $f_0(q,a)$ and found that numerically a very good approximation is offered by $S(z)=1-\mathrm{tanh}(z)$.

\paragraph{Relation to other recent works}~\\
In this section we discuss how our approach relates to and differs from Refs.~\cite{Egana-Ugrinovic:2021gnu,Yunis:2021fgz}, where cosmological evolution of self-interacting WDM was recently studied. 

A crucial novelty of our approach is the exact determination of the non-relativistic distribution function of DM that follows from entropy and number conservation 
during the adiabatic evolution before kinetic decoupling.  This is encoded in the parameter $\alpha$ (see Eqs.~\eqref{f0NR},~\eqref{eq:alphaF} and ~\eqref{eq:alphaB}) that determines the velocity of DM matter at late times, and also when DM becomes effectively non-relativistic. One natural definition is the instance or the scale factor when the temperatures of the relativistic and non-relativistic distributions coincide, $a_*$ in Eq.~\eqref{eq:astar}.

Although not framed in the above manner, Ref.~\cite{Egana-Ugrinovic:2021gnu} has a matching condition that corresponds to  $\alpha=0.1$ for fermions, which 
arises from equating the temperature of the relativistic and non-relativistic regimes at $M/5$. Compared to this reference we have simply modified the publicly available code \texttt{CLASS} 
to include the self-interactions in the Boltzmann hierarchy of non cold DM species. Our power spectrum looks different from the one presented in~\cite{Egana-Ugrinovic:2021gnu} for $k> 20 \,h/$Mpc  showing no significant oscillations. The final bound derived studying deviations from CDM in the region $k < 20\,h/$Mpc is however compatible with their results even though the shape of the exclusion region 
is different.

In \cite{Yunis:2021fgz} several matching conditions for the non-relativistic distribution were considered. 
We believe that the correct choice is the iso-entropic condition one but we do not agree with the numerical value quoted in that
paper, which in our notation would correspond to $\alpha=0.5$. This is probably at the origin of the significant differences between our results. In that paper a modified version of \texttt{CLASS} was presented to study the transient between the relativistic and
non-relativistic regime. While this is interesting on its own, as explained in section~\ref{sec:power}, for practical purposes we 
find it sufficient to consider the evolution with a constant Maxwell-Boltzmann distribution since the relevant modes re-enter the horizon 
when DM is already non-relativistic.
 
\section{Power spectrum of SIWDM}\label{sec:power}

In this section we compute the matter power spectrum resulting from the set of equations \eqref{eq0}-\eqref{eq2}, using approximation for DM collisional terms through Eq.~\eqref{eq:taucoll}. 

While interactions can be added easily as explained in appendix \ref{app:CLASS}, the inclusion of a time dependent $f_0(q,\eta)$ is a task that would require a major modification to the publicly available code \texttt{CLASS}.\footnote{This was studied in \cite{Yunis:2021fgz} where however the derivative with respect to  $\eta$ of the distribution function in Eqs. (\ref{eq0},\ref{eq1},\ref{eq2}) was neglected.} This technical challenge is interesting on its own, but turns out to be unimportant for SIWDM because the relevant   
modes re-enter the horizon after the transition to the non-relativistic regime. This implies that the knowledge of the non-relativistic distribution is sufficient for our purposes.
To see this, recall that $\mathcal{O}$(1) deviations in the power spectrum for SIWDM can only show up at large wave numbers. Focusing on a scale of $k_{\rm max}\approx 10\, h/$Mpc as representative of the smallest scales probed by Lyman-$\alpha$~\cite{Irsic:2017ixq} (and depicted in figure \ref{fig:sketch}), we see that the corresponding mode crosses the horizon, $k_{\rm max}\approx a_{k_{\rm max}} H(a_{k_{\rm max}})$ at around $a_{k_{\rm max}}\approx 5\times 10^{-7}$, while all other smaller wave-numbers re-enter later. 
Since (see Eq. \eqref{eq:astar}) $a_{k_{\rm max}}> a_*$, if $a_{dec} > a_*$ all the dynamics is captured by the non-relativistic distribution.

\begin{figure}[t]
\centering
\includegraphics[width=0.55\textwidth]{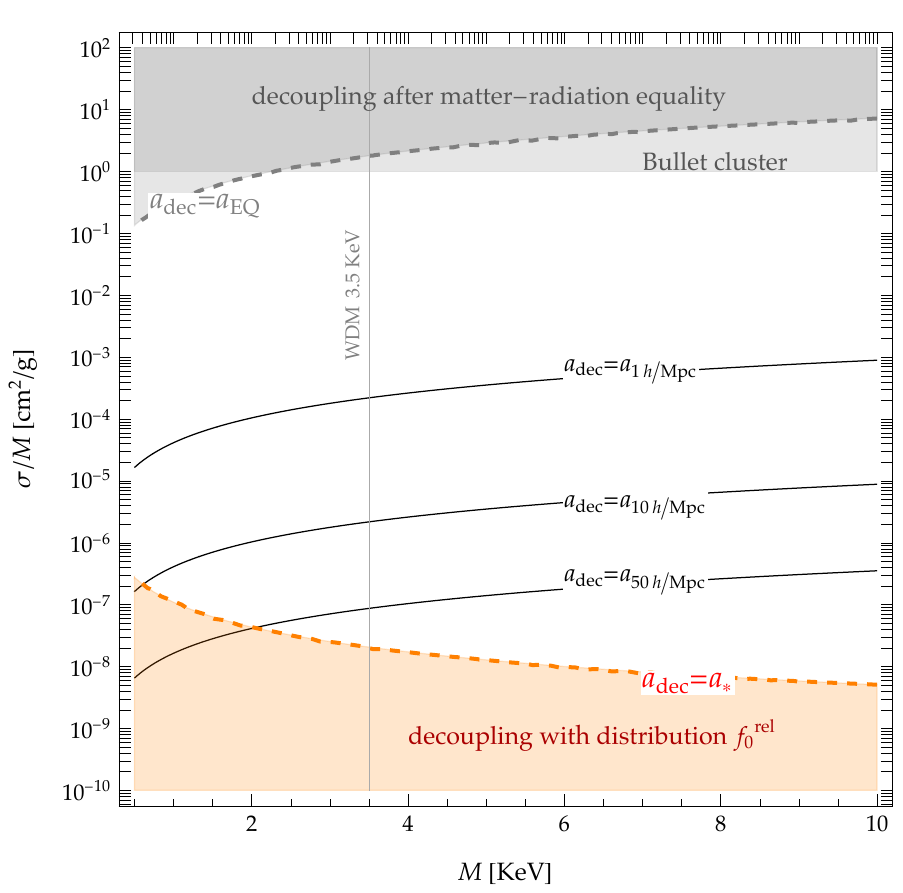}~
\caption{\label{fig:parameterspace}\it Parameter space of self-interacting dark matter in the plane $(M,\sigma/M)$. The light orange region corresponds to DM decoupling while relativistic. In the white region decoupling takes place in the non-relativistic regime so that the distribution function transitions from eq.~\eqref{f0rel} to  eq.~\eqref{f0NR}. We have computed isolines of the scale factor when self-interactions decouple for co-moving momenta $k=[0.5,10,50]\, h/\Mpc$ where $10 h/\Mpc$ can be taken as representative of the largest $k$ probed by Lyman-$\alpha$ data sets. For cross-sections larger than this critical line the transient from relativistic to non-relativistic distribution is not expected to be relevant, see section \ref{sec:power}. At large values of the cross-section, decoupling occurs closer to matter radiation equality, eventually hitting the Bullet cluster constraint. }
\end{figure}

This argument explains how in the relevant range of wave-numbers used to set constraints on WDM, between $k\in [0.5,20]h/\Mpc$,  the sub-horizon evolution of the $k$-modes happens when DM has already transitioned to a non-relativistic distribution. To emphasize this crucial fact we show in figure \ref{fig:parameterspace} the values of the cross-sections and the corresponding decoupling scale factor. It follows that the Boltzmann equations that we need to solve are Eqs.~\eqref{eq0}-\eqref{eq2}, with $f_0$ given by the (time-independent) expression in the deep non-relativistic regime, Eq.~\eqref{f0NR}. 
As a check we have tested that our results indeed do not depend on the transient between relativistic and non-relativistic distributions.

\begin{figure}[t]
\centering
\includegraphics[width=0.45\linewidth]{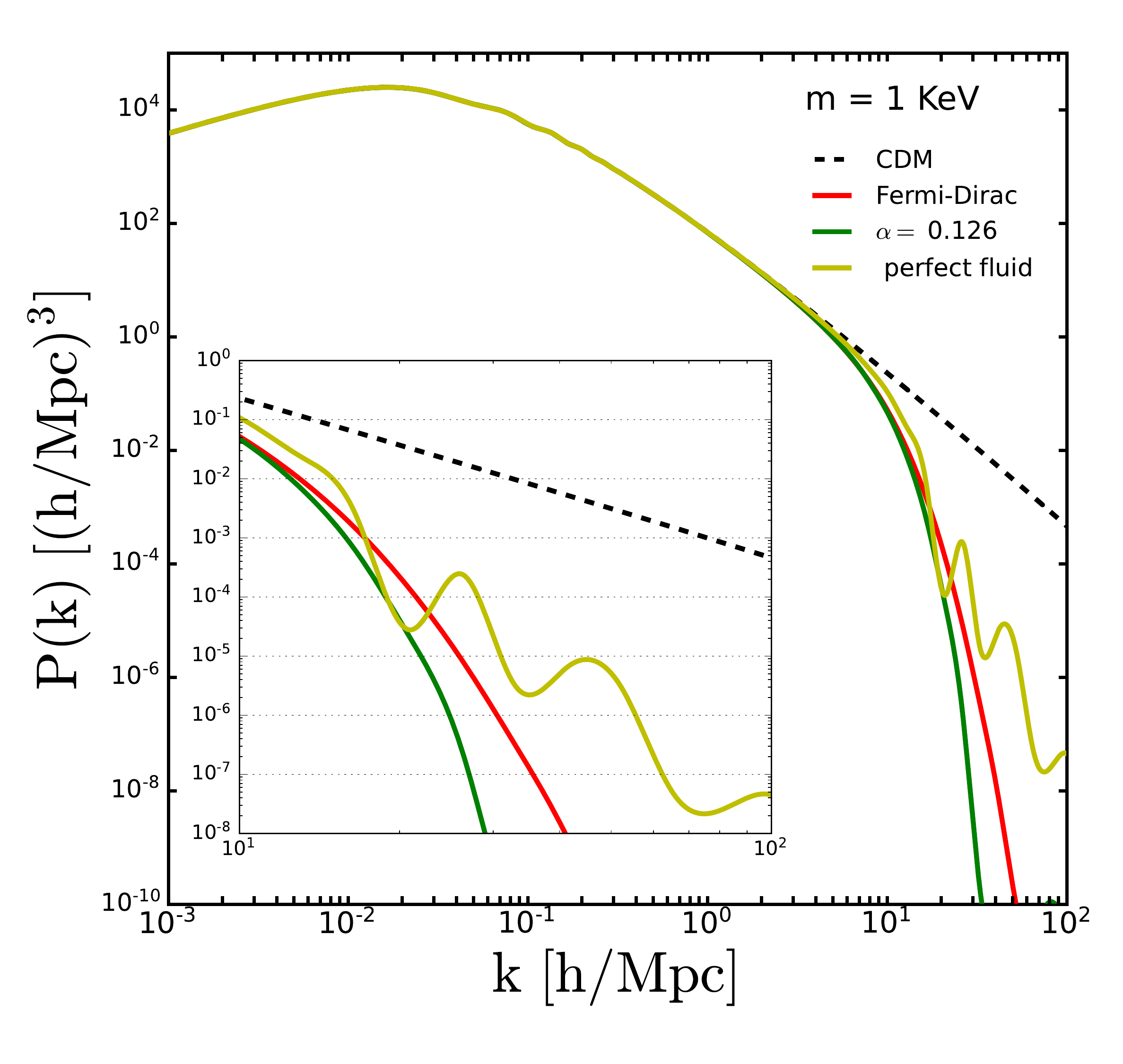}
\includegraphics[width=0.45\linewidth ]{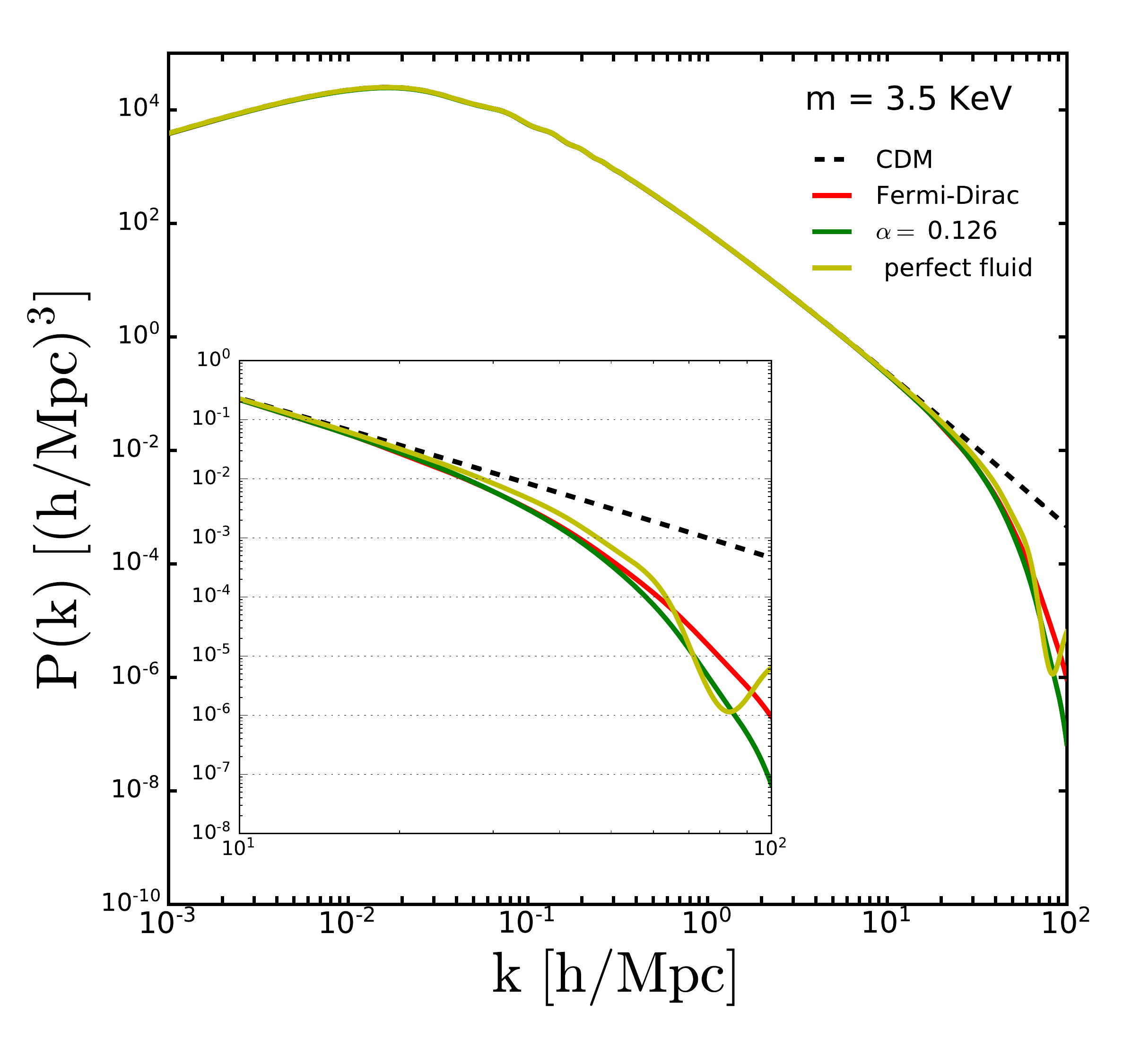}
\caption{\label{fig:power-without-interactions}\it Matter power spectrum $P_{m}(k)$ (linear) for a Weyl fermion with $M=1\, KeV$ (left) and $M=3.5\, \KeV$ (right). We show prediction for various choices of the background distribution $f_0(q)$  relativistic and non-relativistic $f_0\propto e^{-\alpha \tq^2}$, and we compare them with the standard CDM prediction. In both cases we also show the power spectrum of a perfect fluid DM with the same mass and sound speed as in Eq.~\eqref{eq:perfect-matching}.}
\end{figure}

\begin{figure}[t]
\centering
\includegraphics[width=0.45\linewidth]{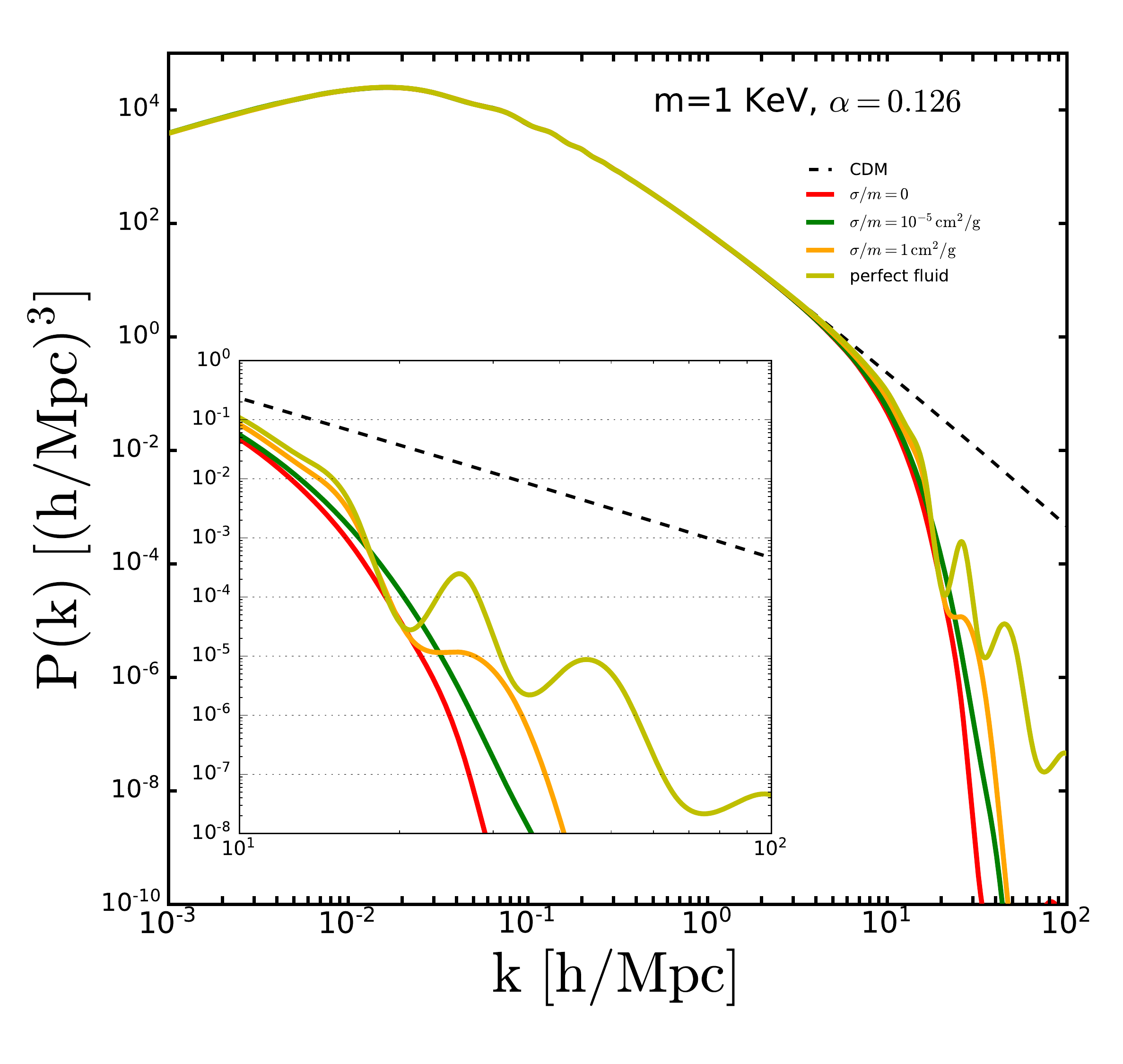}
\includegraphics[width=0.45\linewidth]{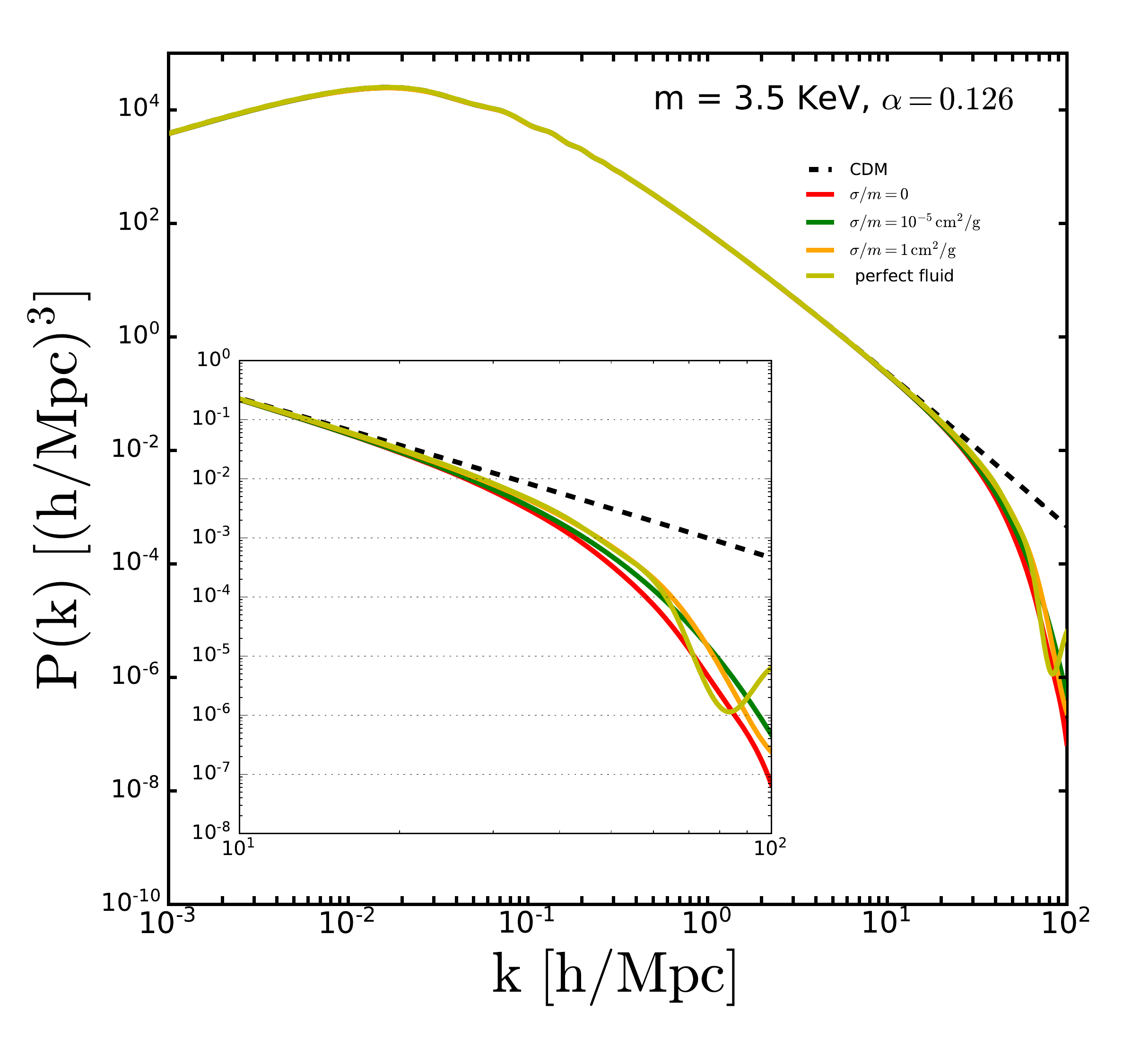}
\caption{\label{fig:power-with-interactions}
\it Matter power spectrum $P_{m}(k)$ (linear) for a Weyl fermion with mass $M=1\, \KeV$ (left) and $M=3.5\, \KeV$ (right). We show prediction for various choices of the background distribution $f_0(q)$,  relativistic and non-relativistic $f_0\propto e^{-\alpha \tq^2}$  and for different values of the cross-section $\sigma/M$. In both cases we also show the power spectrum of a perfect fluid DM with the same mass and sound speed as in Eqs.~\eqref{eq:perfect-matching} and~\eqref{eq:astar}.
}
\end{figure}

\begin{figure}[t]
\centering
\includegraphics[width=0.65\linewidth]{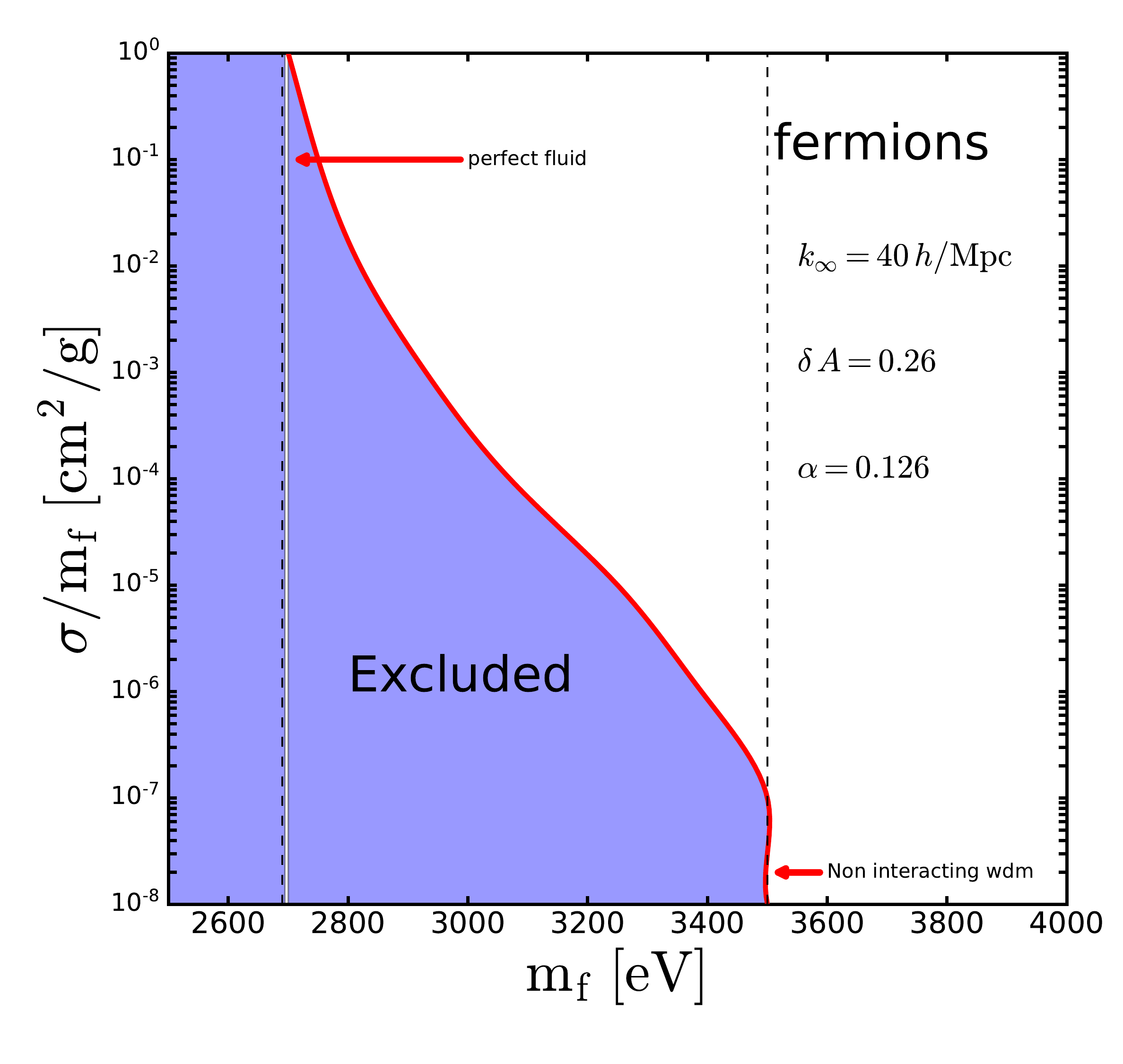}
\caption{\label{fig:results}\it Exclusion limits on Self-Interacting Warm Dark Matter for a Weyl fermion as a function of mass and cross-section. Constraints are derived applying the `area criterion' as discussed in the text. The dashed vertical lines correspond to the limiting cases of a perfect fluid and free-streaming WDM. 
}
\end{figure}

\bigskip
Before discussing SIWDM, we first compute the power spectrum for free-streaming WDM consisting of a Weyl fermion, for a few relevant cases: Fermi-Dirac distribution, Maxwell-Boltzmann distribution with a proper choice of $\alpha=0.126$ as in Eq.~\eqref{eq:alphaF}, and for DM described by a perfect fluid. The results of the computation performed with \texttt{CLASS} are shown in figure~\ref{fig:power-without-interactions} for $M=1\,\KeV$ (left panel) and $M=3.5\,\KeV$ (right panel). We see that, while Fermi-Dirac and Maxwell-Boltzmann are practically indistinguishable, the perfect fluid case gives a larger power for wave numbers where Lyman-$\alpha$ bound applied. Thus the perfect fluid is in closer agreement with the CDM prediction.

Next we compute the  power spectrum in the presence of self-interactions. In figure~\ref{fig:power-with-interactions}, we show the power spectrum for two values of the mass, $M=1\,\KeV$ (left panel) and $M=3.5\,\KeV$ (right panel) for different values of elastic cross-section, and a non-relativistic background distribution with $\alpha = 0.126$ corresponding to a Weyl fermion. With respect to the non-interacting WDM case, interactions increase the power at small scales, becoming closer to $\Lambda$CDM.

The power spectrum for perfect fluid has been computed using the set of continuity and Euler equations (supplemented by the equation of state) as given by Eqs.~\eqref{perfect1}-\eqref{eq:perfect-matching}. We stress again that the perfect fluid case corresponds to the most strongly interacting scenario, and we show it here together with the free-streaming case to emphasize that the realistic scenario of SIWDM will lie in between free streaming and perfect fluid. The latter is specified by $a_*$ that appears in Eq.~\eqref{eq:perfect-matching}. We have implemented in \texttt{CLASS} a routine with a perfect fluid satisfying equations~\eqref{perfect1}-\eqref{eq:perfect-matching}, with this parametrization of the sound speed and equation of state through $a_\star$ (details are provided in appendix~\ref{app:CLASS}). 
\subsection{Bounds on SIWDM}
\label{sec:boundsSIWDM}
We now turn to the constraints on DM with self-interactions. Such bounds arises because of a suppressed matter power spectrum on wave-numbers $k$ that are tested by the observation of Lyman-$\alpha$ absorption. The observed spectra of distant quasars, due to absorption by the intergalactic neutral hydrogen, have been instrumental in recent years to test the properties of DM on scales $k\sim \mathcal{O}(1-100) h/\Mpc$ (see for example \cite{Viel:2013fqw}). Roughly, a reduced power on those scale will induce a reduced observed flux, in contradiction with the observed one.

The high-resolution quasars data-set used to derived the bounds are at redshifts $z\approx 2-6$ in the data-sets MIKE/HIRES+XQ-100 \cite{Irsic:2017ixq} and the extraction of limits and bounds from the linear matter power spectrum is done by following the prescriptions of Ref.~\cite{Murgia:2017lwo}. The observation, already put forward in Ref.~\cite{Schneider:2016uqi}, is that the Lyman-$\alpha$ forest is mostly sensitive to the so-called `flux power spectrum', which is in turn proportional to the one-dimensional matter power spectrum $P_{1D}(k)$ via a bias-function which has little dependence on the underlying DM model \cite{Viel:2005qj}. Therefore, ratios of one-dimensional power spectra are good estimators of the deviation from a given reference model. The one-dimensional power spectrum is defined as
\be
P_{1D}(k)\equiv\int_k^{k_\infty} \frac{dK}{2\pi} \, K P_m(K)\,,
\ee
where the numerical infinity corresponds to $k_\infty\to \infty$.
It is convenient to define the ratio with respect to the CDM case
\be
r(k)\equiv \frac{P_{1D}(k)}{P_{1D, \rm cdm}(k)}\,,\quad\quad \delta A\equiv 1 - \frac{1}{k_{\rm max}-k_{\rm min}}\int_{k_{\rm min}}^{k_{\rm min}}dk r(k)~.
\ee
A possible procedure to set limits, as outlined in \cite{Murgia:2017lwo} and adopted for example in \cite{DEramo:2020gpr,Egana-Ugrinovic:2021gnu,Yunis:2021fgz}, is the following:
\begin{itemize}
\item[ $i)$] Compute the integral  $\int dk\, r(k)$ in the range $k\in[0.5,20]h/\Mpc$, since the most important constraints comes from MIKE/HIRES+XQ-100 data-sets sensitive to those scales; 
\item[$ii)$] Exclude, at 95\% CL, those models for which $\delta A > \delta A(\rm WDM, 3.5 KeV)$, that is we compare the `area' of a given model with the one of WDM at 3.5 KeV, which is the present limit on a free-streaming WDM with Fermi--Dirac distribution as derived from the detailed analysis of Ref.~\cite{Irsic:2017ixq}.\footnote{In the literature some other conditions on $r(k)$, analogous to the ones of this work, were also used, see for example \cite{Schneider:2016uqi}.}
\end{itemize}
 This procedure defines the so-called `area criterion' of Ref.~\cite{Irsic:2017ixq} that we will apply in what follows. Finally note that, the Lymann-$\alpha$ power spectrum depends on the thermal history through re-ionization process. For example, the effect of gas pressure during re-ionization produces the same effect as that of WDM~\cite{Garzilli:2018jqh,Garzilli:2019qki}. In light of this fact, the resulting constraints on WDM could be weaker than what is presented here.

\bigskip

\paragraph{Bounds on fermionic SIWDM:}
Taking as reference 3.5 keV WDM  in our version of \texttt{CLASS} we get
\be
\delta A(\mathrm{WDM}, 3.5 \mathrm{KeV})=0.26\,.
\ee
We then compute $\delta A$ for all our scenarios with the inputs listed in appendix \ref{app:CLASS} and exclude masses the values of parameters where $\delta A>0.26$.

\medskip

The exclusion limits are shown in figure \ref{fig:results}, while a sample of power spectra for different values of the interactions are shown in figure \ref{fig:power-with-interactions}. Assuming that the free streaming candidate is excluded for $3.5$ KeV 
we find that interactions relax the bound to,
\begin{equation}
\boxed{M_{\rm WDM}^{\rm int}> 2.8\, {\rm KeV}}
\end{equation}
compatibly with the bullet cluster constraint. 

\medskip

The results  show that the power spectrum of a perfect fluid deviates less from the CDM prediction than a pure free-streaming WDM does. In section \ref{sec:psd}, we have discussed in detail the relevant parameters for perfect fluids and SIWDM and we have explained the different size of $k_J$ and $k_{\rm fs}$ that mostly control the qualitative behavior of the power spectrum. When DM is semi-relativistic the two parameters differ by a factor of $\sqrt{3}$. Moreover,  we have discussed how the higher-multipoles in the weak-limit $\tau_{\rm coll}\lesssim 1/H_\eta$, tend to suppress the power of the density contrast (and their oscillations) in a way formally analogue to the Silk-damping for the photons (see Eq.~\eqref{eq:damping}). We notice however that since DM is already in the non-relativistic regime when interactions decouple, the effect of damping is not a major one, although it contributes to the weakened limits for SIWDM as compared to pure WDM of figure~\ref{fig:results}.

\begin{figure}[t]
\centering
\includegraphics[width=0.7\textwidth]{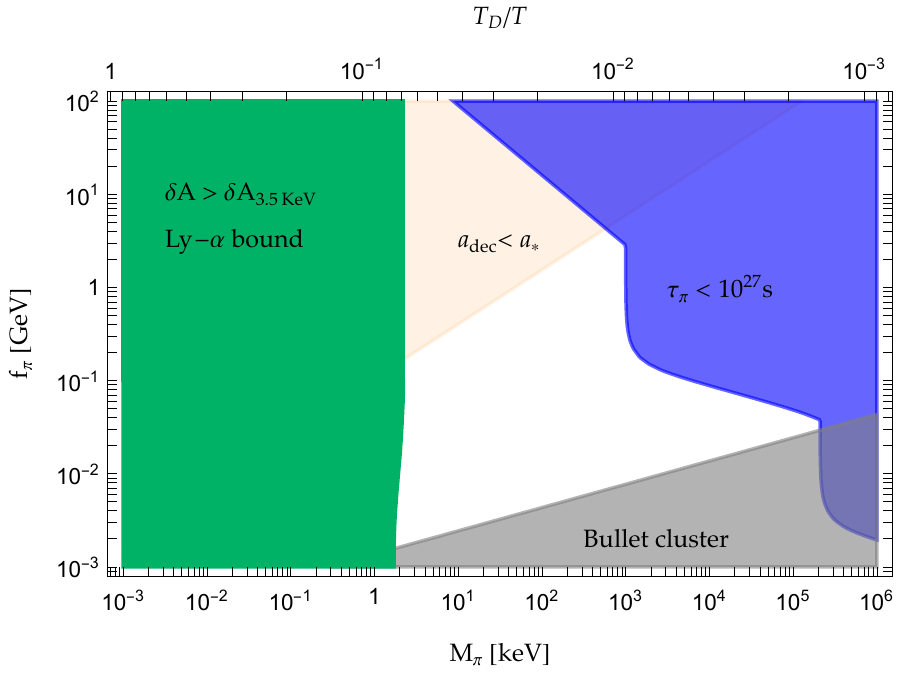}
\caption{\label{fig:darkQCD}\it Parameter space of DM in the dark QCD scenario in the plane mass of DM (dark pion mass, $M_\pi$) and confinement scale $\Lambda \sim f_\pi$. Only in the orange region DM decouples in the non-relativistic regime so that free-streaming bounds are directly applicable. The constraint from the DM relic abundance fixes the temperature ratio $T_D/T$.}
\end{figure}

\section{Dark QCD: results}\label{sec:darkQCD}

The tools developed in the previous sections can be applied to all models where the dark sector experiences self-interactions. They are relevant in particular for secluded dark sectors, with light mass scales, that interact with the SM only gravitationally. 

In absence of interactions we can estimate the lowest value of the mass allowed by Lyman$-\alpha$ measurements by 
requiring that the averaged velocity squared is equal to the one of 3.5 KeV Weyl fermion. 
Using Eq.~\eqref{eq:T0rel} the expected bound reads,
\begin{equation}
M_*^{B(F)}=\frac{3.5(3.9)\, {\rm KeV}}{g_D^{1/4}}~, 
\label{eq:scaling}
\end{equation}
for bosons (fermions).
In the presence self-interactions this bound will be relaxed by about 20\% similar to the case of Weyl fermion studied in section~\ref{sec:boundsSIWDM}.

An explicit realization of self-interacting light DM is offered in particular by  confining gauge theories with fermions, in short dark-QCD \cite{Garani:2021zrr}, that confines at a scale $\Lambda$. Such theories are particularly attractive because dark gauge interactions imply the existence of accidental symmetries that guarantee the cosmological stability of DM
automatically. In the case at hand the lightest dark baryon is stable due to accidental baryon number conservation while the lightest pion is stable being the lightest state 
of the sector. While dark baryons are necessarily heavy to reproduce the DM abundance dark pions can only be DM if their mass is lighter than GeV. In the KeV mass range bounds
from free-streaming are thus expected and were estimated in \cite{Garani:2021zrr}. 

After the confinement phase transition almost all the  energy  in the dark sector is transferred to dark pions that are initially relativistic.
The abundance of  pions in a SU(N) gauge theories with $N_F$ degenerate Dirac flavors $\psi$ is given by eq. (\ref{eq:YDM}) with $g_D=N_F^2-1$, the number of scalars.
The ratio of dark and visible temperature, $T_D/T\equiv \Trel/T_{0,\rm CMB}$, is determined by the production mechanism of the dark sector in the Early Universe well before confinement and it is evaluated as in Eq.~\eqref{eq:YDM}. Since the sector is completely decoupled from the SM, one possible  mechanism is through freeze-in from graviton exchange \cite{Redi:2020ffc} or from higher dimensional operators such as $|H|^2\bar\psi\psi/\Lambda_{\rm UV}$, leading to dark sectors much colder than the SM. Here since we discuss the confined phase, we take $T_D/T$ as a free parameter, allowing also scenarios with $T_D/T\lesssim 1$. 

We focus for simplicity on $N_F=3$ degenerate flavors in what follows. DM is in this case made of 8 degenerate dark pions with mass $M_\pi^2\sim 4\pi m_\psi f$.
When the mass of pions is below the MeV scale the model allows for $\pi$-DM only for relatively small $\Lambda$, while dark baryons comprise only a fraction of DM. The parameter space relevant for the discussion of self-interacting WDM is shown in figure \ref{fig:darkQCD}. The relevant bounds in this context can be summarized by computing the: $i)$ $\pi\pi\to\pi\pi$ scattering cross-section, which affects the structures in clusters (notably the Bullet Cluster); $ii)$ bounds from Lyman-$\alpha$ inferred from the (linear) power spectrum for light pions with self-interactions. In both case cases an important quantity is the elastic cross-section of pions. In dark QCD with three colors, this is given by
\be
\sigma_{\pi\pi}=\frac{77}{1536\pi}\frac{M_\pi^2}{f^4}\,,
\ee
with $f\sim \Lambda$ being the pion decay constant.

From Eq.~\eqref{eq:adec} one finds that DM remains self-coupled in the NR regime if,
\begin{equation}
f< {\rm KeV} \left(\frac{M_\pi}{\rm KeV}\right)^{7/12}~.
\end{equation}
This corresponds to the red region in Fig. \ref{fig:darkQCD}.

We have repeated the analysis in section \ref{sec:SIWDM} for the dark pion scenario. 
We determine the allowed region of parameter space requiring that deviation from CDM $\delta A$ is the same as 3.5 KeV free streaming Weyl fermion, $\delta A< 0.26$.
The difference arises from the different multiplicity, $g_D=8$ and the non-relativistic distribution $\alpha =.155$ that implies a reduced velocity than the fermionic case. Consequently, for the same given DM mass, this means that the pressure is smaller and slightly more power could be expected with respect to the fermionic case.

\begin{figure}[t]
\centering
\includegraphics[width=0.65\linewidth]{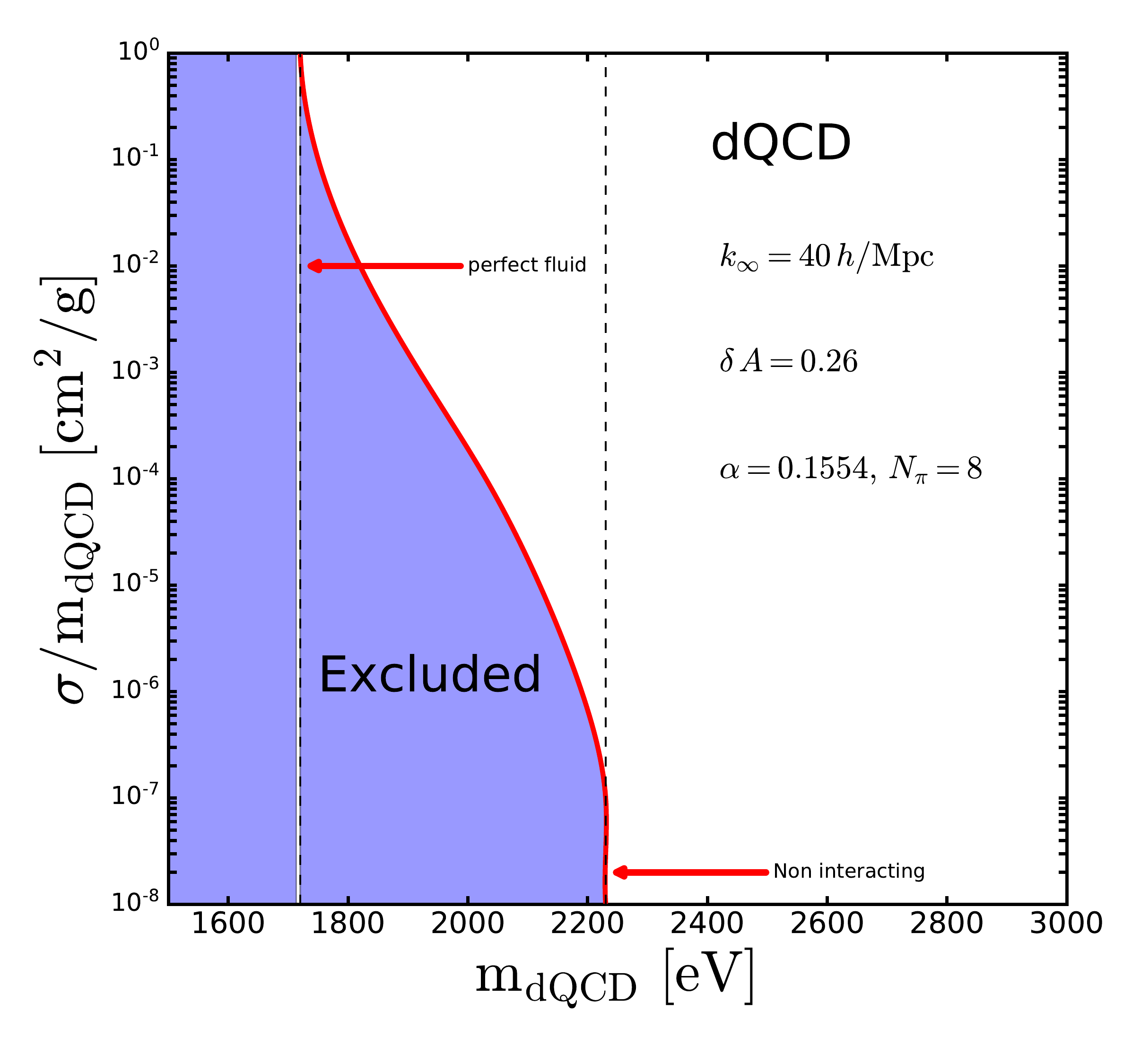}
\caption{\label{fig:dQCD}\it Exclusion limits on dark pion DM as a function of mass and cross-section. Constraints are derived applying the `area criterion' as discussed in the text. The dashed vertical lines correspond to the limiting cases of a perfect fluid and free-streaming non-interacting DM. We assume 8 degenerate scalars. }
\end{figure}

As shown in figure~\ref{fig:dQCD} Lyman-$\alpha$ constraints give the lower bound on the mass,
\begin{equation}
\boxed{M_{\rm dQCD}> 1.7(2.2)\, {\rm KeV}}
\end{equation}
where the lower (higher) bound applies to self-interacting (free-streaming) regimes.

\section{Conclusions}\label{sec:conclusion}

Dark sectors with just gravitational couplings to the SM, despite their feeble interactions with the visible sector, can affect the structures we see in our Universe, modifying in particular the matter power spectrum as compared to the $\Lambda$CDM case. In this work we have derived updated constraints on Warm Dark Matter with self-interactions, a scenario that is naturally realized in dark sectors with confining gauge theories where the lightest states act as an accidental DM candidate. 

Interactions in the dark sector play a crucial role by keeping the system in kinetic equilibrium while it becomes non-relativistic. We find that, even for tiny couplings, the modes of interest for the matter power spectrum re-enter the horizon when DM has already become non-relativistic. Therefore generically the background phase space distribution of DM, $f_0$, is of Maxwell-Boltzmann type; the dependence on the initial distribution is only encoded in the overall normalization factor  and in the non-relativistic temperature, both fixed by number and entropy conservation. We have derived exact matching conditions for fermions and bosons  produced relativistically (extending the work of ref.~\cite{Egana-Ugrinovic:2021gnu,Yunis:2021fgz}). 
Interactions also affect the actual structure of the Boltzmann equations for DM perturbations, that we have here re-obtained in the so-called relaxation time approximation: 
a hierarchical structure of moments of the Boltzmann equations with interactions.

Our analysis shows that the presence of interactions allow for smaller masses than the purely free streaming case, although marginally. This milder bound is ascribed both to the background distribution function which is of Maxwell-Boltzmann type for all the relevant modes, and to the interactions themselves for the perturbations, which make them collapse on slightly larger scales as discussed in section \ref{sec:equations}. In order to establish the robustness of our findings we have also considered the case of perfect fluid DM, with the same relativistic-to-non-relativistic transition as an example of maximal strength of interactions. This case corresponds to the most allowed scenario (for fixed mass and initial temperature) and it is also very stable numerically, offering a fast and reliable way to test our results. 

The study of cosmological implications of self-interacting dark sectors offers a unique chance to understand the nature of DM in the plausible scenario where it has only gravitational couplings with the SM. Other directions are possible along these lines, for example  DM with exotic equation of state or a sub-component of DM, as well as the presence of interacting dark radiation in dark sectors without a mass gap. Possible future extensions of our work include the  development of a dedicated module for SIWDM in \texttt{CLASS}, allowing for a time dependence in the background distribution function $f_0$ and in the Boltzmann equations, as in Eqs.~\eqref{eq0}-\eqref{eq2}, as well as for momentum-dependent cross sections. We hope to return these and other questions in future work.

{\small
\subsubsection*{Acknowledgments}
We thank Thejs Brinckmann, Francesco D'Eramo, Deanna Hooper and Julien Lesgourgues for discussions related to \texttt{CLASS}.
This work is supported by MIUR grants PRIN 2017FMJFMW and 2017L5W2PT and the INFN grant STRONG. 
We acknowledge the Galileo Galilei Institute for hospitality during part of this work. AT also thanks LPTHE in Paris for hospitality during the completion of this work. }

\appendix

\section{Inputs and modifications in  \texttt{CLASS}}\label{app:CLASS}

We have modified parts of the \texttt{cdm} and \texttt{ncdm} routine in \texttt{CLASS}. The most relevant modifications to the code are the following:

\begin{itemize}
\item In order to allow for different users' inputs of time-independent $f_0(q)$ as discussed in the original paper \cite{CLASSIV}, we have followed \cite{DEramo:2020gpr} and we have changed in \texttt{source/background.c} the default Fermi-Dirac distribution to a generic function
\be
f_0(\tq)= A_0 (\exp{A_1\tq} + \exp{A_2 \tq^2} + A_3)^{-1},
\ee
which encompasses all the cases discussed in this paper by proper selections of the coefficients $A_i$.
\item The inclusion of self-interactions in the Boltzmann hierarchy of Eqs.~\eqref{eq0}-\eqref{eq2} has been done modifying the \texttt{ncdm} routine, inside \texttt{source/perturbations.c}, adding the possibility to add to the variable \texttt{dy[idx+l]}, for $\ell\geq 2$, the quantity 
\be
\mathtt{-s\_l[l]*y[idx+l]*taucoinv}
\ee
where \texttt{taucoinv} is the inverse of the (comoving) collision time $\tau_{\rm coll}$ of eq.~\eqref{eq:taucoll} and has been defined accordingly in the code inside the \texttt{ncdm} routine,
~\\
\texttt{\scriptsize
	\textcolor{violet}{double} taucoinv;~\\
   if( pba->has\_wdminteraction == \_TRUE\_)\{ ~\\ 
        		if (a <= a\_star) \{ taucoinv=6.19*(1e-6)*sigmaovermass/a2;  /* units are Mpc/h */ \} ~\\
        		else \{taucoinv=6.19*(1e-6)*sigmaovermass*a\_star/(a*a2);  /* units are Mpc/h */ \}~\\
       	\}~\\
        else \{taucoinv=0.0;\}
}

\item The inclusion of the perfect fluid has been done by duplicating the routine of \texttt{cdm} and labeling the new routine with a flag added to the structure \texttt{pba}. The new routine is defined by two new equations for Newtonian gauge  (our Eqs.~\eqref{perfect1}-\eqref{perfect2})
~\\~\\
\texttt{\scriptsize
if (pba->has\_cdm == \_TRUE\_ \&\& pba->has\_perfectfluid == \_TRUE\_) \{~\\~\\
\textcolor{violet}{double} csquare, wdark, dwdark;~\\
        if (a <= a\_star) \{ csquare=1./3.; wdark=1./3.; dwdark=0.;\}~\\
        else \{ csquare=(1./3.)*pow(a\_star,2.)/a2; wdark=3./5.*csquare;dwdark=-6./5.*csquare*a\_prime\_over\_a;\}~\\~\\
if (ppt->gauge == newtonian)\{
~\\~\\
 dy[pv->index\_pt\_delta\_cdm] = -(1+wdark)*(y[pv->index\_pt\_theta\_cdm]+metric\_continuity)~\\
-3*a\_prime\_over\_a*(csquare-wdark)*y[pv->index\_pt\_delta\_cdm]; /* perfect fluid density */~\\~\\
dy[pv->index\_pt\_theta\_cdm] = - a\_prime\_over\_a*(1-3*wdark)*y[pv->index\_pt\_theta\_cdm]~\\
        - dwdark/(1+wdark)*y[pv->index\_pt\_theta\_cdm]+ csquare/(1+wdark)*k*k*y[pv->index\_pt\_delta\_cdm]~\\
       + metric\_euler; /* perfect fluid velocity */~\\
       \}~\\~\\
       ...~\\
       ...~\\
       \}
}

where the new functions \texttt{wdark}, \texttt{dwdark} and \texttt{csquare} are such to match with eq.~\eqref{eq:perfect-matching}, and the variable \texttt{a\_star} of eq.~\eqref{eq:astar} is passed externally to the code for any given mass and cross-section input.
\end{itemize}

For the runs of \texttt{CLASS} we have selected the amplitude $A_s=2.1\times 10^{-9}$ and tilt $n_s=0.9660499$ of the primordial scalar fluctuations during inflation, and we have used $\Omega_b h^2=0.0224$. When computing the Boltzmann hierarchy for SIWDM, in order to increase precision we have switched off all the approximations corresponding to $\mathtt{ncdm\_fluid\_approximation=ncdmfa\_off}$. We have run the Boltzmann hierarchy with $\ell_{\rm max}=50$. Furthermore, we use the following for the precision variables for our runs: $\mathtt{tol\_ncdm\_bg}= \mathtt{tol\_thermo\_integration}= \mathtt{tol\_ncdm}= \mathtt{tol\_ncdm\_newtonian}= 10^{-11}$, and $\mathtt{tol\_perturbations\_integration}=10^{-6}$.

\pagestyle{plain}
\bibliographystyle{jhep}
\small
\bibliography{biblio}

\end{document}